# POINT-OF-CARE DIABETIC RETINOPATHY DIAGNOSIS: A STANDALONE MOBILE APPLICATION APPROACH

*A dissertation submitted in partial fulfilment of the requirement of degree of*

**MASTER OF TECHNOLOGY**

In

**COMPUTER SCIENCE AND ENGINEERING**

**(With specialization in Data Science)**

*Submitted*

*By*

**MISGINA TSIGHE HAGOS**

**2017015646**

*Under the guidance of*

**Prof. Shri Kant**

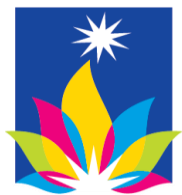

**DEPARTMENT OF COMPUTER SCIENCE AND ENGINEERING**

**SHARDA UNIVERSITY, GREATER NOIDA**

**JULY-2019**

# POINT-OF-CARE DIABETIC RETINOPATHY DIAGNOSIS: A STANDALONE MOBILE APPLICATION APPROACH

*A dissertation submitted in partial fulfilment of the requirement of degree of*

**MASTER OF TECHNOLOGY**

In

**COMPUTER SCIENCE AND ENGINEERING**

**(With specialization in Data Science)**

*Submitted*

*By*

**MISGINA TSIGHE HAGOS**

**2017015646**

*Under the guidance of*

**Prof. Shri Kant**

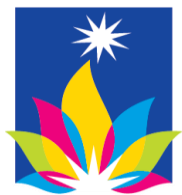

**DEPARTMENT OF COMPUTER SCIENCE AND ENGINEERING**

**SHARDA UNIVERSITY, GREATER NOIDA**

**JULY-2019**

# DECLARATION

I hereby declare that this dissertation report titled "**Point-of-Care Diabetic Retinopathy Diagnosis: A Standalone Mobile Application Approach**" is my own work and that, to the best of my knowledge and belief, it contains no material previously published or written by another person nor material which has been accepted for the award of any other degree or diploma of the university or other institute of higher learning, except where due acknowledgment has been made in the text.

Place: Sharda University                    Signature of the candidate:

Date: _________________                Name: Misgina Tsighe Hagos

                                                                 Email: tsighemisgina@gmail.com or

                                                                2017015646.misigana@pg.sharda.ac.in

Reg.No: ___________________

Date of Registration: _________________

# CERTIFICATE

I certify that this dissertation report titled "**Point-of-Care Diabetic Retinopathy Diagnosis: A Standalone Mobile Application Approach**" is the bona fide work of Mr. Misgina Tsighe Hagos who carried out dissertation work under my supervision. Certified further, that to the best of my knowledge the work reported herein does not form part of any other project report or dissertation on the basis of which a degree or award was conferred on an earlier occasion on this or any other candidate.

Signature:                                            Signature:

Professor Shri Kant                                   Dr. Nitin Rakesh

RTDC, Professor, CSE                                  Professor, CSE

Supervisor                                            Head of Department (HOD)

Signature:

Dr. Mandeep Kaur

Associate Professor, CSE

Program Coordinator

Place: Sharda University

Date: _______________

# ABSTRACT

Although deep learning research and applications have grown rapidly over the past decade, it has shown limitation in healthcare applications and its reachability to people in remote areas. One of the challenges of incorporating deep learning in medical data classification or prediction is the shortage of annotated training data in the healthcare industry. Medical data sharing privacy issues and limited patient population size can be stated as some of the reasons for training data insufficiency in healthcare. Methods to exploit deep learning applications in healthcare have been proposed and implemented in this dissertation.

Traditional diagnosis of diabetic retinopathy requires trained ophthalmologists and expensive imaging equipment to reach healthcare centres in order to provide facilities for treatment of preventable blindness. Diabetic people residing in remote areas with shortage of healthcare services and ophthalmologists usually fail to get periodical diagnosis of diabetic retinopathy thereby facing the probability of vision loss or impairment. Deep learning and mobile application development have been integrated in this dissertation to provide an easy to use point-of-care smartphone based diagnosis of diabetic retinopathy. In order to solve the challenge of shortage of healthcare centres and trained ophthalmologists, the standalone diagnostic service was built so as to be operated by a non-expert without an internet connection. This approach could be transferred to other areas of medical image classification.

# ACKNOWLEDGMENT

I would like to express my utmost gratitude to my parents: Baba and Mama, and to my beloved siblings: Girmay, Muler and Trew.

I am grateful for the advises and directions I received from my advisor Professor Shri Kant. I also acknowledge my colleague Surraya Ado Balo's (PhD candidate) support in reviewing related works.

Last but not least, I am thankful to the evolutionary process that has resulted in me having the capabilities to propose and work on this dissertation work.

# TABLE OF CONTENTS

# LIST OF TABLES

# LIST OF FIGURES

# LIST OF ACRONYMS

| | |
|---|---|
| **ACC** | Accuracy |
| **AMD** | Age-Related Macular Degeneration |
| **AUC** | Area under the ROC curve |
| **AI** | Artificial Intelligence |
| **ANN** | Artificial Neural Networks |
| **CSME** | Clinically Significant Macular Edema |
| **CPM** | Competition Metric |
| **CNN** | Convolutional Neural Networks |
| **CDR** | Cup-to-Disc Ratio |
| **DBN** | Deep Belief Networks |
| **DME** | Diabetic Macular Edema |
| **DR** | Diabetic Retinopathy |
| **DWT** | Discrete Wavelet Transform |
| **EXs** | Exudates |
| **FN** | False Negative |
| **FP** | False Positive |
| **FCRN** | Fully Convolutional Residual Network |
| **HMs** | Haemorrhages |
| **HEF** | Hand Engineered Features |
| **HEs** | Hard Exudates |
| **IDE** | Integrated Development Environment |
| **ICDR** | International Clinical Diabetic Retinopathy Severity Scale |
| **IOU** | Intersection Over Union |
| **LED** | Light Emitting Diode |
| **MAs** | Microaneurysms |
| **NPDR** | Non-Proliferative Diabetic Retinopathy |
| **OC** | Optic Cup |
| **OD** | Optic Disk |
| **PDR** | Proliferative Diabetic Retinopathy |
| **ROC** | Receiver Operating Characteristic Curve |
| **ReLu** | Rectified Linear Unit |
| **ResNet** | Residual Neural Network |
| **SN** | Sensitivity |
| **SEs** | Soft Exudates |
| **SP** | Specificity |
| **SGD** | Stochastic Gradient Descent |
| **SVM** | Support Vector Machines |
| **TN** | True Negative |
| **TP** | True Positive |
| **VTDR** | Vision Threatening Diabetic Retinopathy |

# 1. INTRODUCTION

In this chapter, anatomy of the retina, diabetes mellitus, and the prevalence and progression of Diabetic Retinopathy (DR) will be discussed. An overview of signs of Diabetic Macular Edema (DME) and its effects on vision will be presented. Different medical image recognition methods, and the challenge of training data insufficiency in applications of deep learning in healthcare; the development environments used to implement a DR detection system and future of DR automated diagnosis systems will also be introduced. The contribution of this dissertation work and its structure will be presented at the end of the chapter.

## 1.1 INTRODUCTION TO DIABETIC RETINOPATHY

### 1.1.1 ANATOMY OF THE RETINA

Retina is a light-sensitive layer that lines the inside of the back of the eyeball. It consists of several layers of neuronal cells that are interconnected by synapses. It is organized into three layers, the photoreceptive layer, the bipolar cell layer, and the ganglion cell layer. The photoreceptive layer consists of rods, which provide black and white vision, and cones, which are responsible for color vision. The rods and cones synapse with the bipolar cells in the second layer of the retina. Bipolar cells communicate with both the first and third layers. The axons of the ganglion cells found in the third layer convey the visual information as encoded by the retina to the next synapse point in the visual pathway via the optic nerve.

As can be seen in Figure 1.1, the reddish circular structure is the macula. Macula is the structure where light is focused from cornea and lens. While the rest of the retina provides peripheral vision, macula enables the human eye to see in great detail. The macula needs to stay dry for a clear vision. Fovea is the central part of the macula.

Optic disk or the optic nerve head is part of the retina where the optic nerve exits the eye. It is the point of exit ganglion cell axons leaving the eye. Optic disk is a blind spot of the eye since it does not contain any light-sensitive cells. There are no cons or rods staying on top of the optic disk. It is the white part in Figure 1.1. Optic nerve transfers visual information from the retina to the brain via electric impulses. Atrophy of the optic nerve occurs because of the high pressure in the vitreous fluid inside the eye. Optic nerve gets compressed because of the high pressure and this causes cells to die.

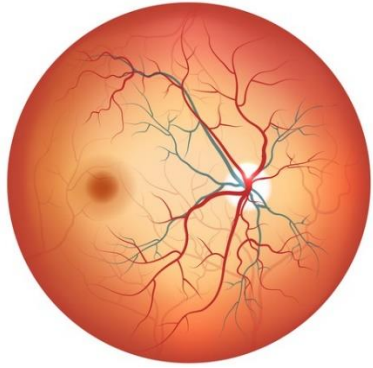

Figure 1.1. The Retina

Diabetic retinopathy, retinal tear or detachment, solar retinopathy, hypertensive retinopathy, central serous retinopathy, and age related macular degeneration are some of the common retinal disorders.

### 1.1.2 DIABETES MELLITUS

Diabetes Mellitus, commonly known as Diabetes is a condition where the pancreas produces less or no insulin. Insulin helps to transport glucose or sugar into human cells. It acts as a key to open the human cells so glucose can enter inside. Without insulin, unused glucose builds up around cells in the blood. This build-up of unused glucose overflows into the urine and passes out of the body. There are two general types of diabetes: type 1 and type 2. In type 1 diabetes the body's autoimmune system destroys insulin producing cells in the pancreas. And in type 2 diabetes, either the body isn't producing enough insulin or the produced insulin isn't working properly.

Diabetes damages the human body's process of circulating, storing and effectively using sugar or glucose. Since diabetes hinders the absorption of sugar by cells, there will be excessive sugar in the blood, which reduces the energy gained by the body. With delayed treatment, diabetes will affect the blood vessels of the retina to overflow with blood. This damaging effect of diabetes mellitus on retinal blood vessels in the retina causes vision loss and impairment.

In 2014, there were 422 million people diagnosed with diabetes [1]. In 2010, lesions of DR were seen in more than one third of the patients diagnosed with diabetes. A third of the diabetic patients diagnosed with DR were also diagnosed with vision threatening diabetic retinopathy.

Figure 1.2 shows the percentage of projected increase of the number of people with diagnosed diabetes due to population size increase, prevalence rate growth, and demographic changes [2]. The total number of people with diabetes will rise from 11 million in 2000 to almost 20 million in 2025. By 2050, this is projected at 29 million people which is a 165% increase over the year 2000 level [2].

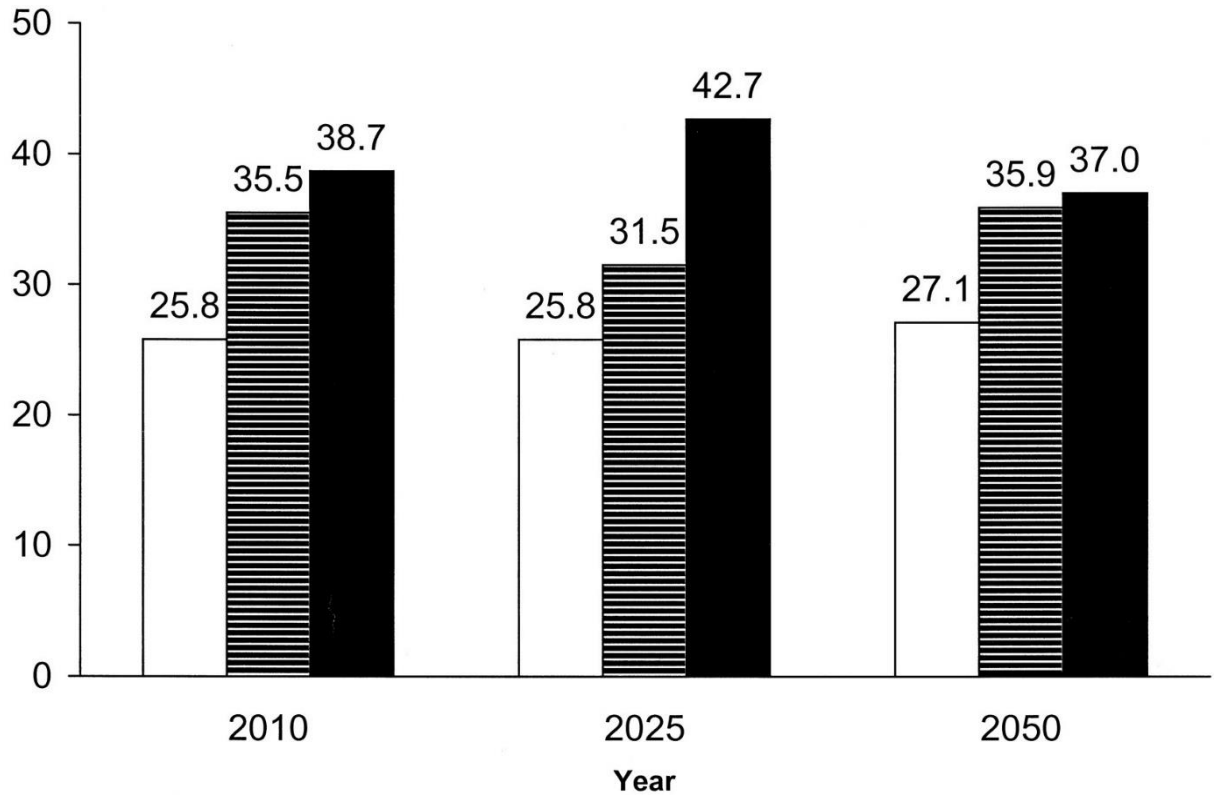

Figure 1.2. Percentage breakdowns of the projected increase in the total number of people with diagnosed diabetes due to population size increase, prevalence rate growth, and demographic changes. ☐ Population growth; ☰ increasing prevalence rates; ■ demographic changes. [2]

### 1.1.3 DIABETIC RETINOPATHY

Diabetic retinopathy is one of the major causes of vision impairment around the world. Every diabetic person is at risk of developing diabetic retinopathy. According to the World Health Organization, there were 422 million diabetic patients in 2014, 35% of whom developed some type of diabetic retinopathy because of increased and constant damage to retinal small blood vessels. The World Health Organization reported that approximately 1 in 3 people living with diabetes are diagnosed with some degree of diabetic retinopathy, and 1 in 10 will develop a vision threatening diabetic retinopathy [1]. Globally in 2010, out of overall 32.4 million blind and 191 million visually impaired people, 0.8 million were blind and 3.7 million were visually impaired because of diabetic retinopathy, with an alarming increase of 27% and 64%, respectively, spanning the two decades from 1990 to 2010 [3].

In the United States of America, the number of diabetic retinopathy patients is expected to double between 2010 and 2050, as can be seen in Figure 1.3 [4].

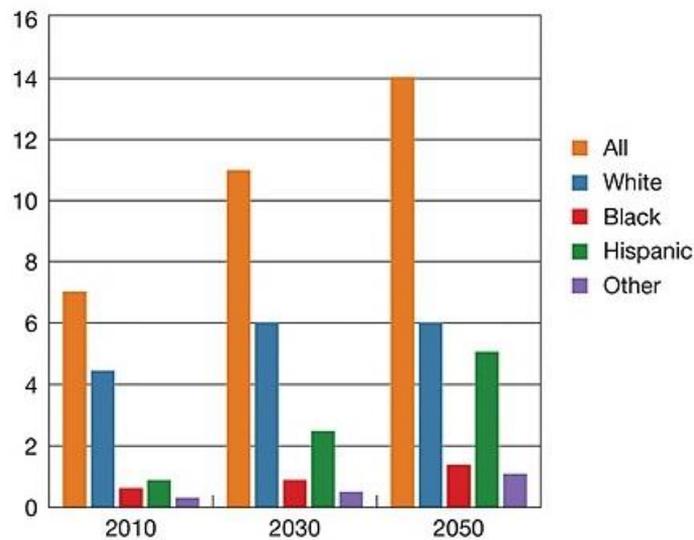

Figure 1.3. Projection of diabetic retinopathy in United States of America in millions [4]

Diabetic retinopathy is the leading cause of vision loss in working age adults (20- 65 years.) It is a complication caused because of constant damage to retinal blood vessels caused by the accumulation of glucose around the retina.

Early detection of diabetic retinopathy can help with enabling treatment of the retina and thereby avoiding vision loss and long term vision impairment. Diabetic retinopathy is traditionally diagnosed by professional ophthalmologists by taking retinal pictures and studying for disease signs on the collected images. During diabetic retinopathy diagnosis, an ophthalmologist will scan for abnormal blood vessels, blood, swelling or fatty deposits in the retina, retinal detachment and abnormalities in the optic nerve, and growth of new blood vessels and scar tissue from previous studies. In addition, an ophthalmologist may test vision and look for evidence of cataracts when testing for diabetic retinopathy.

As shown in Figure 1.4, based on disease severity scale, a consensus was reached by Wilkinson et al. [5] to label diabetic retinopathy into 5 stages. In the case where diabetes mellitus has not caused any damage to the retina, it is labelled as "no apparent retinopathy" which is the first scale. The second stage is "mild diabetic retinopathy" in which few microaneurysms start to appear. Cotton wool spots, dot and blot haemorrhages, and multiple microaneurysms are the main signs of the third scale of diabetic retinopathy which is identified as "moderate diabetic retinopathy". "Severe diabetic retinopathy" is the fourth label of the disease and its main characteristics are: cotton wool spots, venous beading, and intraretinal microvascular abnormalities. Retinal detachment, vitreous haemorrhages, and production of new retinal blood vessels are the characteristics of the last stage, which is termed as "proliferative diabetic retinopathy."

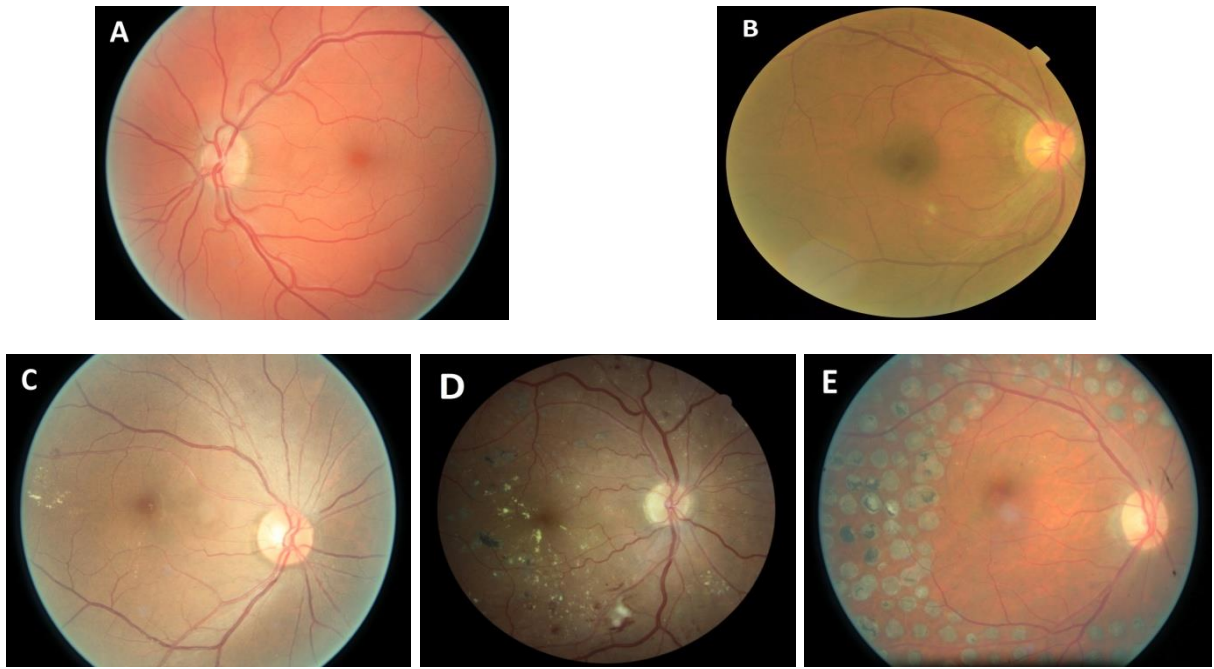

Figure 1.4. Stages of diabetic retinopathy: (A) No apparent retinopathy; (B) Mild diabetic retinopathy (C) Moderate diabetic retinopathy (D) Severe diabetic retinopathy (E) Proliferative diabetic retinopathy

### 1.1.4 TYPES OF DIABETIC RETINOPATHY LESIONS

Diabetic retinopathy lesions can be observed, identified and classified by trained professionals such as ophthalmologists for expert based traditional diagnosis of the disease diabetic retinopathy. The lesions can be observed after taking image of the retina employing one of optical coherence tomography, fluorescein angiography, b-scan ultrasonography, or color fundus photography as retinal imaging techniques [6].

Microaneurysms are seen in the starting periods of DR and retinal damage. Unusual escape of blood from retinal blood vessels around macula results in microaneurysms which are small in size. They appear sharp-edged as red spots. After time, greater size of microaneurysms having irregular shapes will be spotted. These are called haemorrhages or dot-and-blot because of their unusual margins. Haemorrhages are caused when walls of weak capillaries get broken. Exudates start to form because of increased blood escape from capillaries.

Exudates usually appear yellow in colour and irregular shaped on the retina. The two types of exudates are hard and soft exudates. Hard Exudates are proteins and other nutrients leaking out from broken retinal blood vessels with sharp margins. They appear white or white-yellow. Hard Exudates are located in the outer layer of the retina organized in blocks or circular rings. Soft exudates or cotton wool spots are small, cloud-like shaped with whitish-grey colour. They are caused by the blockage of the arteriole. The arteriole connects arteries with the capillaries. Arteriole is used for oxygen, nutrients and waste exchange. Soft exudates are caused by the accumulation of axoplasmic material within the nerve fibre layer. Exudates are different from microaneurysms and haemorrhages in degree of brightness. While microaneurysms and haemorrhages appear as dark lesions, exudates appear bright.

### 1.1.5 AUTOMATIC DIABETIC RETINOPATHY DETECTON

Diabetic retinopathy becomes incurable at later or advanced stages. Early treatment of diabetic retinopathy can help avoid permanent vision loss or impairment caused by the disease. DR is traditionally diagnosed by taking retinal pictures and studying for disease signs on the collected images. This diagnosis process requires expensive fundus imaging devices installation in healthcare centers and trained professionals. Imaging techniques usually used in diabetic retinopathy diagnosis are optical coherence tomography, fluorescein angiography, b-scan ultrasonography, and color fundus photography [6]. All of these techniques come with the challenge of expensive development, deployment, and usage costs. Trained professional are required to employ, and take advantage of these techniques. In addition to the trained professional need of these techniques an ophthalmologist or more are required to study and diagnose a fundus image that is captured by the imaging methodology. An ophthalmologist usually requires two to seven days for retinal image screening. Diabetes patients that reside in rural and remote areas usually suffer from delayed diagnosis of DR because of the expensive deployment of diagnosing equipment, and shortage of ophthalmologists and health care centers. This leads to losing access to early treatment. Automatic diagnosis of fundus images for diabetic retinopathy identification and classification helps to fight the challenges of traditional diabetic retinopathy diagnosis.

Traditional machine learning techniques often make use of Hand Engineered Features (HEF) in automated identification of diabetic retinopathy. Mookiah et al. [7], grouped automated identification of DR based on employed methods and techniques. Diabetic retinopathy features such as microaneurysms, exudates, hemorrhages, and macula edema have been used for feature based detection, and classification. Expert knowledge is necessary for HEF based classification. HEF based automatic diagnosis requires selecting the most applicable features. Detailed comparison and sorting of candidate features is also required. It has been well observed that HEF based classification methods fail to generalize well.

Deep learning based medical image classification has gained attention of researchers and developers because of the increasing processing power offered by sharable graphical processing units and huge datasets. Deep learning based classification have been showing high performance gain over the hand engineering features classification based identification of diseases. Various deep learning models and training algorithms have also been proposed to implement automatic computer aided diagnosis of diabetic retinopathy from fundus images. Due to the shortage of ophthalmologists, and healthcare centers around the world, mobile diagnosis service of diabetic retinopathy have also been implemented and researched.

In image level classification of diabetic retinopathy, deep learning techniques such as convolutional neural networks and deep belief neural networks have been used. Convolutional neural networks can be designed and used starting with initialization, which is called end-to-end training. Transfer learning can also be employed for faster training and reduced training dataset. Transfer learning is the applicability of already trained neural network models in classifying previously unseen dataset. Transfer learning can be useful in medical image classification because the healthcare sector of automated image classification suffers from annotated training data insufficiency.

### 1.1.6 FUTURE OF DIABETIC RETINOPATHY DIAGNOSIS

Diabetic retinopathy diagnosis is clinically performed by ophthalmologists with the help of high end fundus images capturing devices. Optical coherence tomography, fluorescein angiography, B-scan Ultrasonography, and color fundus photography are some of the most frequently used retinal imaging techniques employed in diabetic retinopathy diagnosis [6]. All of these techniques come with the challenge of expensive development, deployment, and usage costs. Trained professional are required to employ, and take advantage of these techniques. In addition to the trained professional need of these techniques an ophthalmologist or more are required to study and diagnose a fundus image that is captured by the imaging methodology. An ophthalmologist usually requires two to seven days for retinal image screening. Diabetes patients that reside in rural and remote areas usually suffer from delayed diagnosis of DR because of the expensive deployment of diagnosing equipment, and shortage of ophthalmologists and health care centers. This leads to losing access to early treatment.

Foster and Resnikoff [8] put four strategies to fight the challenges the diagnoses process of diabetic retinopathy faces in order to implement treatment for preventable blindness; (1) Creating academic, public and governmental awareness of the effects of blindness and visual loss, and the fact that 75% diseases that cause blindness are preventable; (2) Automating and mobilizing existing techniques and methods; (3) Implementing district-specific and country-specific prioritizing strategies of diagnosing and treatment resources for a productive process; (4) Providing comprehensive, maintainable and fair diagnosis services of visual diseases at district level, which includes stuff training, distributing diagnosis and treatment resources, and infrastructure, such as health care center buildings.

As is put by Foster and Resnikoff [8], in order to automate and mobilize existing techniques and methods, a smart phone device with mobile retinal image capturing cameras could be used to diagnose diabetic retinopathy in remote and rural areas. This can solve the cost and time challenges of the traditional diagnosis of diabetic retinopathy.

One of the main challenges of deploying diabetic retinopathy automatic diagnosis on a smartphone is the quality of the retinal images captured and the limited processing power and memory inside. A speedy, easy to use and cheap retinal imaging device could be used as an add-on with a smartphone to collect fundus images of patients as in [9]. Kim et al. [9] proposed the CellScope retina with a custom software to be installed on a smart phone. There are five stages until the production of a wide-field retinal image: (1) A user by holding the device in hand points it at the retina of a patient. (2) In built light emitting diode is used to illuminate the retina. (3) A green dot is used as a fixation target. (4) Each individual image is approximately $50^0$. With the current automated arrangement, five overlapping images are captured of the central, superior, nasal, temporal, and inferior retina. (5) A custom software is used to merge the collected images into one and create a wide-field image of the retina spanning approximately $100^0$.

Smartphone based automatic diagnosis of diabetic retinopathy can be implemented in two ways. The first and easy way is an internet based diagnosis; and the second one is a stand-alone independent smartphone application software based diagnosis. In an internet based diagnosis, smartphone is only used to capture a fundus image of a patient with the help of an add-on retinal camera. After a fundus image is captured, it is sent to a professional ophthalmologist for diagnosis, and results will be sent be over the internet. This would require the user to have a stable internet connection for sending the fundus images, and receiving back the results of diagnosis. The challenge of low internet coverage can be solved using an independent smartphone application that is able to process a captured fundus image and automatically diagnose the stage of diabetic retinopathy without any network connection. A stand-alone smartphone application software could enable patients living in rural and remote areas to get automatic diagnosis of diabetic retinopathy easily.

## 1.2 CONTRIBUTION

The contribution of this dissertation can be summarized as follows:

1. Fundus images dataset that are used in the literature were extensively reviewed, compared and presented.
2. Fundus images preprocessing techniques for noise removal were implemented to prepare a deep learning training dataset.
3. Experimental approach was taken in designing and implementing a transfer learning based deep learning model to detect diabetic retinopathy from reduced training dataset. On account of image preprocessing, transfer learning and model parameters selection, an approach was implemented to solve the challenge of annotated training data insufficiency. This approach could be of insight to other medical image classification researchers.
4. A one-versus-one binary classifier based approach was proposed and implemented with convolutional neural networks to classify diabetic retinopathy into five stages.
5. A standalone smartphone based diagnosis of diabetic retinopathy was implemented and reported. This technique can be used in rural and remote areas with no internet connection and without the need for professional ophthalmologist.

## 1.3 DISSERTATION STRUCTURE

The dissertation report is organized into six chapters. Chapter two provides a review of deep learning techniques, their training algorithms and development environments. Chapter three introduces and compares the fundus image datasets used in research and development works, and the performance measures usually used in diabetic retinopathy classification and detection works. Chapter four provides a detailed literature review of diabetic retinopathy identification, detection and classification. Previous feature extraction based such as macula edema, exudates, microaneurysms, and hemorrhages, image-level based, and smartphone based diabetic retinopathy identification, detection and classification literature is reviewed and compared in chapter four. Chapter five discusses the proposed methodology and its implementation of diabetic retinopathy classification starting from dataset collection and preparation till deep learning based classification; integration of standalone smartphone based diagnostic application software is also discussed in chapter five. Lastly, in chapter seven the conclusion of the dissertation work is presented and recommendations for further improvements are discussed.

## 2. DEEP LEARNING

Deep learning is a computer automation process that is based on artificial neural networks. It belongs to a broader family of machine learning. Artificial neural networks are a systematic ordering of artificial neurons that are inspired by biological neurons found in the human body. They have many differences from biological neurons, a few of which are while the natural neurons are dynamic and analog artificial neurons or nodes are static and symbolic. Deep learning is a collection of architectures that make use of the concept of artificial neural networks. Deep neural networks, deep belief networks, recurrent neural networks, and convolutional neural networks are some of the architectures that employ deep learning. Deep learning has been applied to speech recognition, face recognition, computer vision, facial expression recognition, machine translation, material inspection, drug design, driverless cars, social network filtering, and natural language recognition.

Deep learning architectures, specifically convolutional neural networks, build layers of convolutional layers with classification layers at the end. Layers of convolution operations at the start of the network extract low level input features. And the layers at the end extract high level features. For example, in a character recognition application, the first layers may extract and identify pixels and edges. The last layers will extract and identify a whole character. The classification layer builds fully connected layers that accept features extracted by convolutional layers and classify them into the number of classes.

### 2.1. CONVOLUTIONAL NEURAL NETWORKS

LeCun et al. [10] first introduced the method of applying convolution or convolution with a small size kernel, to image recognition. It was performed for handwritten digit recognition from input image [10] Weight sharing, in order to reduce the number of parameters, and local convolution were also performed to subsequent hidden layers to convolve and extract features of increasing complexity and abstraction [10]

The organization of convolutional neural networks was inspired by the architecture of the visual cortex. Visual cortex contains neurons that individually respond to stimuli only in a restricted region of the visual field which is called the Receptive Field. The visual area is covered by a collection of overlapping visual fields.

A convolutional neural networks is a deep learning algorithm which performs convolution and classification operation using sliding convolution filters on input data. The convolution filters contains sharable weights and biases assigned to different parts of input in order to differentiate between them. The preprocessing required in convolutional neural networks is much smaller compared to other classification algorithms.

While feed forward neural networks ignore pixel dependencies by flattening input image or data, a convolutional neural network will take an image as it is and span across it in order to capture class differentiating features. A convolutional neural network will be able to detect and extract spatial and temporal dependencies in an image through the application of relevant filters. The design of convolutional neural network performs a better fitting to the image dataset because of the reduction in the number of variables and parameters involved, and weights and biases reusability.

There are two basic modules that perform operations inside a convolutional neural network based classification. The first module is a convolutional module, and the second is classification. Image features such as edges extraction from input data, or more specifically input image is the main goal of a convolution operation. Convolution operation doesn't need to be limited to only one layer of convolution. Conventionally, the first convolution operation layer is responsible for extracting the low-level features such as edges, color and gradient orientation. Similar to how humans view the world, with more layers, the architecture of convolution operation layers adapt to extracting high-level features from images, producing a network with a combined comprehension of images in a given dataset.

The three types of layers usually found inside a convolutional neural network are convolution, pooling and fully connected layers as can be seen in Figure 2.1. A convolution layer makes use of convolution kernels in order to extract a specific size of features from images. Size of the convolved features accepted from the convolutional layers is reduced using a pooling layer. This helps to reduce size of data to decrease the processing power required in classification, and to detect and extract rotation and position invariant dominant features and ignore unnecessary features. Two types of pooling can be employed. A maximum pooling collects the maximum value of a specific convolution kernel, and average pooling calculates and extracts average of the values contained in convolved features. After an input image is passed through multiple convolution and pooling layers, a fully connected layer is used to classify the extracted features into different classes. A flattening layer will help to accept the convolution layer's output and forward it in a column format to the next fully connected layer. The number of layers and nodes per layer in fully connected layers will depend on the specific classification or prediction operation and size of the dataset.

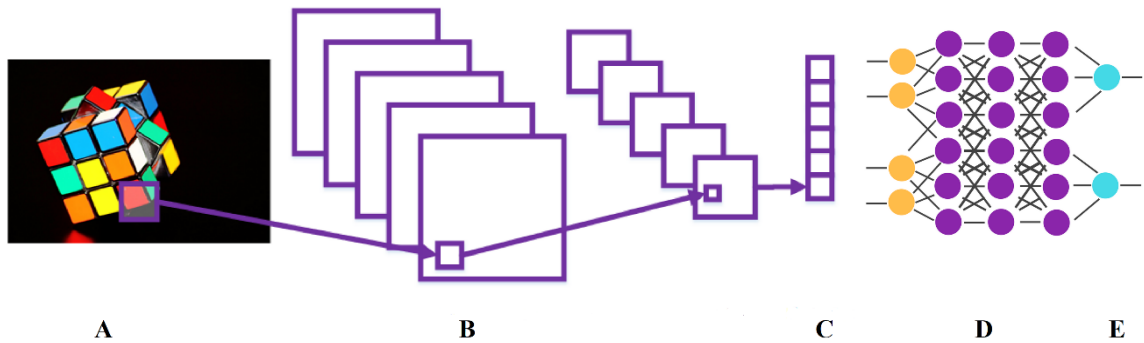

Figure 2.1. Typical convolutional neural network: (A) Input layer; (B) Convolution and pooling layers; (C) Flattening layer; (D) Fully connected layer; (E) Output layer

Convolutional neural networks can be trained and used from scratch, which employs end-to-end training, or transfer learning can be employed for faster training and reduced training dataset. Transfer learning is the applicability of already trained neural network models in classifying previously unseen dataset. Transfer learning can be useful in medical image classification because the healthcare sector of automated image classification suffers from annotated training data insufficiency.

One of the main issues with incorporating deep learning in medical image analyses is the shortage of available labelled training dataset [11, 12]. Transfer learning techniques have gained wider acceptance due to the unavailability of labelled training data in the design and training of deep convolutional neural network models [13, 14]. In [15], annotated training data insufficiency was identified as the main challenge of applying deep learning models in the healthcare automation industry. Furthermore, Altaf et al. [15] recommended that methods, such as transfer learning, that exploit deep learning using reduced data need to be devised and implemented.

## 2.2. RECURRENT NEURAL NETWORKS

Recurrent neural network are a type of artificial neural networks where the previous steps' output are given as input to the current step. Inputs and outputs are usually independent of each other in traditional artificial neural networks, but in cases like when it is required to predict the next part of an input sequence previous parts are required and hence there is a need to store and remember previous parts. As can be seen in Figure 2.2, by making use of a hidden layer that accept feedback, recurrent neural networks solved the issue of computationally remembering historical inputs. Hidden state, which remembers historical information about input sequence, is the most useful part of recurrent neural networks.

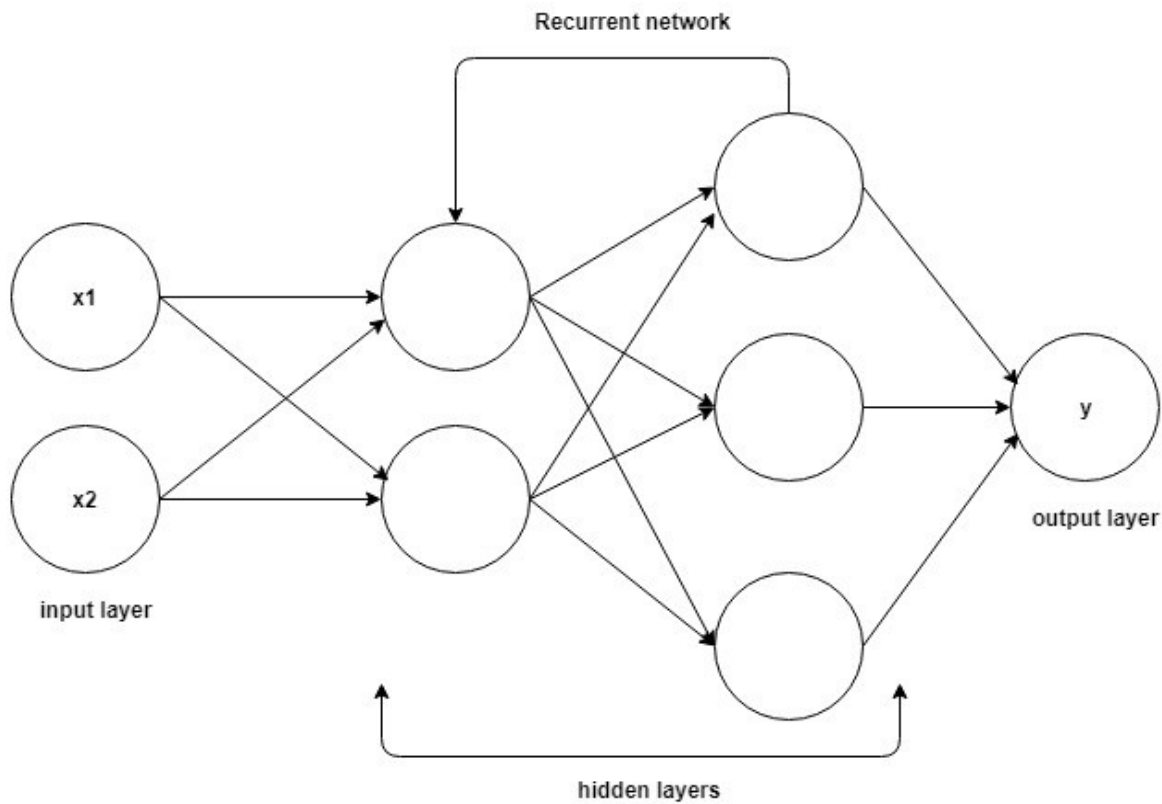

Figure 2.2. Architecture of a basic recurrent neural network

Recurrent neural networks differ from feed forward neural networks because of the incorporation of internal memory or hidden state.

## 2.3. TRAINING METHODS

In order for a deep learning model to be able to generalize a given input dataset into classification or prediction, the model needs to be trained. Training a model includes initializing a model to initial parameters setting, which is initialization of weights and biases, modifying the parameters in iterative process of learning, computing its loss or error at each iteration and iterating further until a satisfying values of parameters is reached that generalizes well on input dataset.

### 2.3.1. END-TO-END LEARNING

End-to-end learning is a type of neural network training style in which the whole network is initialized, modified through a selected training algorithm and validated accordingly. In this process of fitting a model to a dataset, a deep learning model is trained from scratch.

The first step of an end-to-end model training is model initialization. In model initialization, the trainable parameters such as weights and biases are initialized to random values according to a selected initialization method. Model optimization algorithms will be implemented to train the model after initialization. In training process, after calculating the output of the network a loss function is used to compute error of the model. The model's parameters are modified according to the computed loss. A loss function, which takes weights and biases as inputs, is defined to calculate the capacity of the network to approximate the correct ground truth classes for all training input dataset. For example, the loss could be the number of correctly classified fundus images. An algorithm similar to stochastic gradient descent is used to efficiently find the network parameters such as weights and biases by taking the output of the loss function as input.

For each sample from the dataset the prediction and associated loss is calculated. All the associated loss are summed up to produce the total error of the model. Then a back-propagation algorithm is used to propagate the error in order to calculate the partial derivatives of the cost function to for the weights and biases accordingly.

After all the partial derivatives for weights and biases are calculated, the weights and biases are modified using a selected optimization algorithm such as stochastic gradient descent. The process of iteration of prediction of computing labels, back-propagation of loss or error, and optimization of parameters is then performed until the lowest possible loss is reached depending on a pre-set threshold.

### 2.3.2. TRANSFER LEARNING

Transfer learning is the applicability of already trained neural network models in classifying previously unseen dataset. Transfer learning can be useful in medical image classification because the healthcare sector of automated image classification suffers from annotated training data insufficiency. Transfer learning techniques have gained wider acceptance because of the unavailability of labelled training data in the design and training of deep convolutional neural network models [13, 14].

A list of available pre-trained models is shown in Table 2.1. Top-1 accuracy is the probability of the model predicting the correct label with highest probability result. Top-5 accuracy is the performance of the model in predicting the correct label in one of the top 5 highest probability results. Depth refers to the structural depth of the architecture of the model. Depth includes batch normalization layers, convolution layers, activation layers, and pooling layers. All of the available pre-trained models were trained on subset of the ImageNet dataset [16], which contains a thousand classes and around one million and 200 thousand training images.

Table 2.1 List of pre-trained deep learning models

| Model name | Top-1 accuracy | Top-5 accuracy |
| --- | --- | --- |
| Xception [17] | 0.790 | 0.945 |
| VGG16 [18] | 0.713 | 0.901 |
| VGG19 [18] | 0.713 | 0.900 |
| ResNet50 [19] | 0.749 | 0.921 |
| ResNet101 [19] | 0.764 | 0.928 |
| ResNet152 [19] | 0.766 | 0.931 |
| ResNet50V2 [20] | 0.760 | 0.930 |
| ResNet101V2 [20] | 0.772 | 0.938 |
| ResNet152V2 [20] | 0.780 | 0.942 |
| ResNeXt50 [21] | 0.777 | 0.938 |
| ResNeXt101 [21] | 0.787 | 0.943 |
| InceptionV3 [22] | 0.779 | 0.937 |
| InceptionResNetV2 [23] | 0.803 | 0.953 |
| MobileNet [24] | 0.704 | 0.895 |
| MobileNetV2 [25] | 0.713 | 0.901 |
| DenseNet121 [26] | 0.750 | 0.923 |
| DenseNet169 [26] | 0.762 | 0.932 |
| DenseNet201 [26] | 0.773 | 0.936 |
| NASNetMobile [27] | 0.744 | 0.919 |
| NASNetLarge [27] | 0.825 | 0.960 |

Transfer learning is generally performed in two steps. The first part is feature extraction; and the second part is fine-tuning. In feature extraction, a pre-trained model is imported. This pre-trained model is usually trained on a huge representative dataset so that it can be used in smaller dataset by employing transfer learning. After pre-trained model is imported the target dataset is passed through it and the outputs are computed, and the results are stored as features. Feature extraction is performed using convolutional part of convolutional neural networks.

The classifier part of imported pre-trained models is usually trained from scratch to classify the extracted features. This helps to adapt the features to the new dataset.

### 2.3.3. OPTIMIZATION ALGORITHMS

The iterative behavior of deep learning makes the training process resource-intensive. In order to quickly complete the iterative cycle with the availability of different methods to try and number of parameters to tune, it is necessary to be able to train models fast. This is key to increasing the speed and efficiency of a machine learning process. This is why optimization algorithms such as min-batch gradient descent, stochastic gradient descent, Adam optimizer, and gradient descent with momentum are required.

These optimization techniques make it possible for an artificial neural network to fit a classifier to a given dataset and learn. In terms of speed and performance, some methods perform better than others.

Stochastic gradient descent is the most commonly used optimization algorithm for machine learning in general and for deep learning in particular. It follows an approach of computing the gradient of existing parameters with respect to the loss function.

### 2.3.4. LOSS FUNCTIONS

Loss functions evaluate the performance of the set of parameters such as weights and biases at every iteration process of a neural network training. Loss functions are also called costs. In the training process of a deep learning model, the main aim of an optimization algorithm is to reduce the loss function. The mostly used loss functions are mean square error, cross entropy, and loss multi label.

Mean square error is a multi-class loss function used to train an artificial neural network by evaluating outputs on every stage. The formula for calculating mean square error is shown in Equation (2.1), where x is vector of n predictions, and y is the set of expected classes in binary values.

$$Loss(x, y) = \frac{1}{n} \sum_{i} (x_i - y_i)^2 \qquad (2.1)$$

Cross entropy has been experimented with and resulted in a better performance, and it is used for multi-class loss. Its formula can be seen in Equation (2.2), where x is vector of n predictions, and y is the set of expected classes in binary values.

$$Loss(x, y) = -\sum_{i}(y_i * \log\left(\frac{e^{x_i}}{\sum_j(e^{x_j})}\right)) \tag{2.2}$$

The mean square error usually slows down training, while the cross entropy loss function has been observed to be faster.

The cross entropy can be adapted to a multi-label classification, not to be confused with multi-class classification. Multi-label classification is a classification task where a single input item could belong to more than one classes. Loss multi-label loss function is created as seen in Equation (2.3) by adapting the cross entropy loss function.

$$Loss(x, y) = -\sum_{i}(y_i * \log\left(\frac{e^{x_i}}{1 + e^{x_i}}\right) + (1 - y_i) * \log\left(\frac{1}{1 + e^{x_i}}\right)) \tag{2.3}$$

## 2.4. DEVELOPMENT ENVIRONMENTS

### 2.4.1 SPYDER

Spyder is an integrated development environment built for Python development with Python. It can be downloaded as part of the Anaconda distribution. As can be seen in Figure 2.3, its main building blocks are a text editor, IPython console, variable and file explorer, profiler and debugger. Its abilities can be extended with the help of plugins.

In the case of deep learning, Spyder can be used for data preparation and preprocessing, and model building, training and testing.

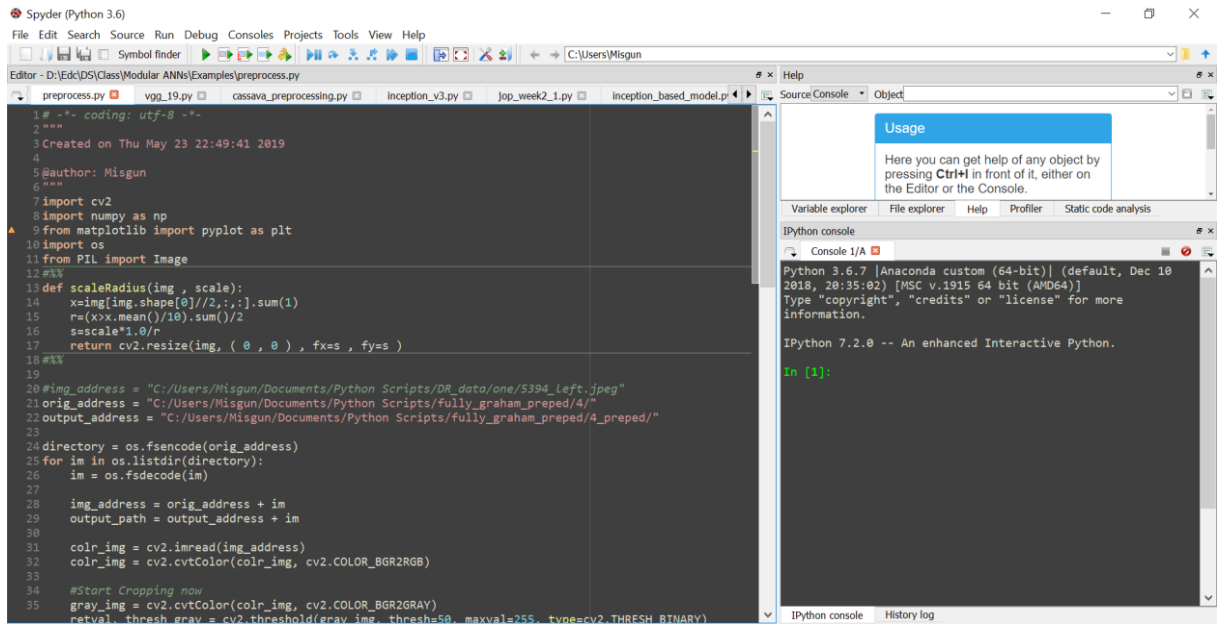

Figure 2.3. Spyder IDE

### 2.4.2 GOOGLE COLABORATORY

Google Colaboratory, also known as Colab, is a computing cloud service based on Jupyter notebooks for machine learning and deep learning deployment. It is a project with the objective of disseminating machine learning education and research [28]. It is a free-of-charge service provided by Google for researchers, students and practitioners. A free computational power is made accessible to a robust GPU and RAM on the cloud. The only requirement to use the service is a Google account.

Colab provides a computational power of GPU and RAM for twelve hours in a single period. This property makes it a reasonable choice for deep learning models training and testing. Its infrastructure is hosted on the Google Cloud platform.

Colaboratory notebooks, seen in Figure 2.4, are based on the Jupyter notebook and act as a Google Docs object. They can be shared with multiple users and users can work on the same notebook. They can be downloaded and uploaded into and from a physical hard drive. Colab provides Python 2 or Python 3 preconfigured with the essential machine learning and deep learning tools such as Tensorflow, Keras and Matplotlib.

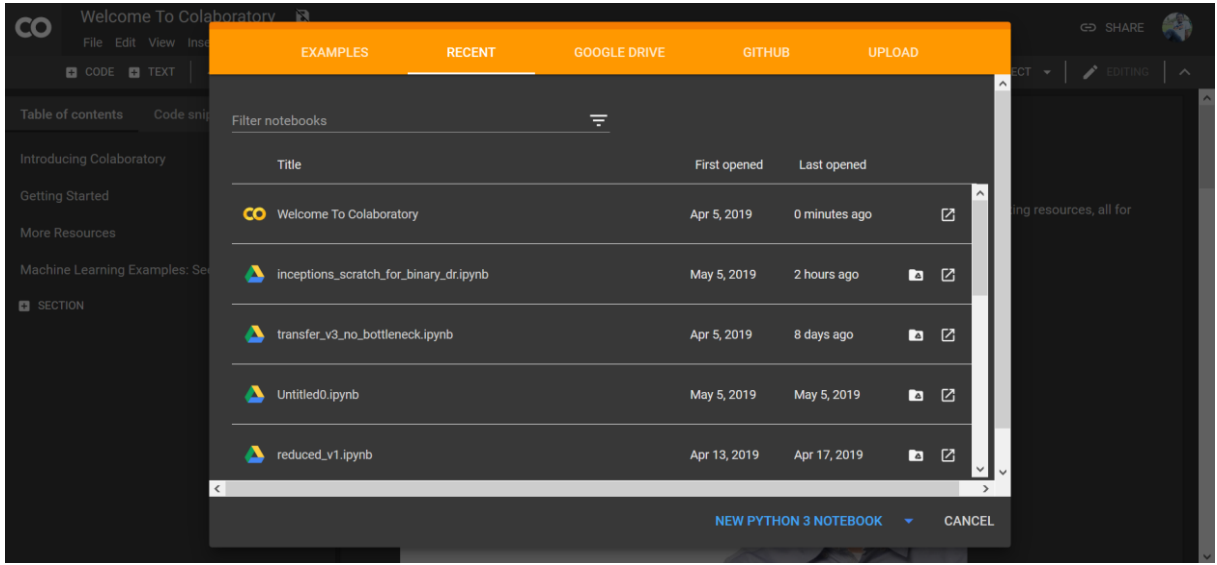

Figure 2.4. Google Colaboratory welcome screen

A study done in [29] showed that Google Colaboratory resources can exhibit a performance similar to dedicated hardware. Carneiro [29] also showed that Colab has a limitation regarding high performance computing because of time limit of GPU utilization, the need of data transfer and its limitations to/from Google Drive or Git.

### 2.4.3 ANDROID STUDIO

A smartphone based easy to use and inexpensive diagnosis of diabetic retinopathy could be deployed for rural and remote areas where there is deficiency of professional ophthalmologists and medical diagnostic equipment.

With the current wider acceptance of smartphones around the world, incorporating automatic diagnosis and classification of diabetic retinopathy would greatly help those that have less coverage of medical services. Smartphone based automatic diagnosis of diabetic retinopathy can be implemented in two ways. The first and easy way is an internet based diagnosis; and the second one is a stand-alone independent smartphone application software based diagnosis. In an internet based diagnosis, smartphone is only used to capture a fundus image of a patient with the help of an add-on retinal camera. After a fundus image is captured, it is sent to a professional ophthalmologist for diagnosis, and results will be sent be over the internet. This would require the user to have a stable internet connection for sending the fundus images, and receiving back the results of diagnosis. The challenge of low internet coverage can be solved using an independent smartphone application that is able to process a captured fundus image and automatically diagnose the stage of diabetic retinopathy without any network connection.

A stand-alone smartphone application software could enable patients living in rural and remote areas to get automatic diagnosis of diabetic retinopathy easily.

As can be seen in Figure 2.5, for developing, testing and deploying a smartphone application for automatic diagnosis of diabetic retinopathy, whether it is an independent fully automated diagnostic application software or a remotely diagnostic smartphone application, android studio can be used for smartphones with an android operating system.

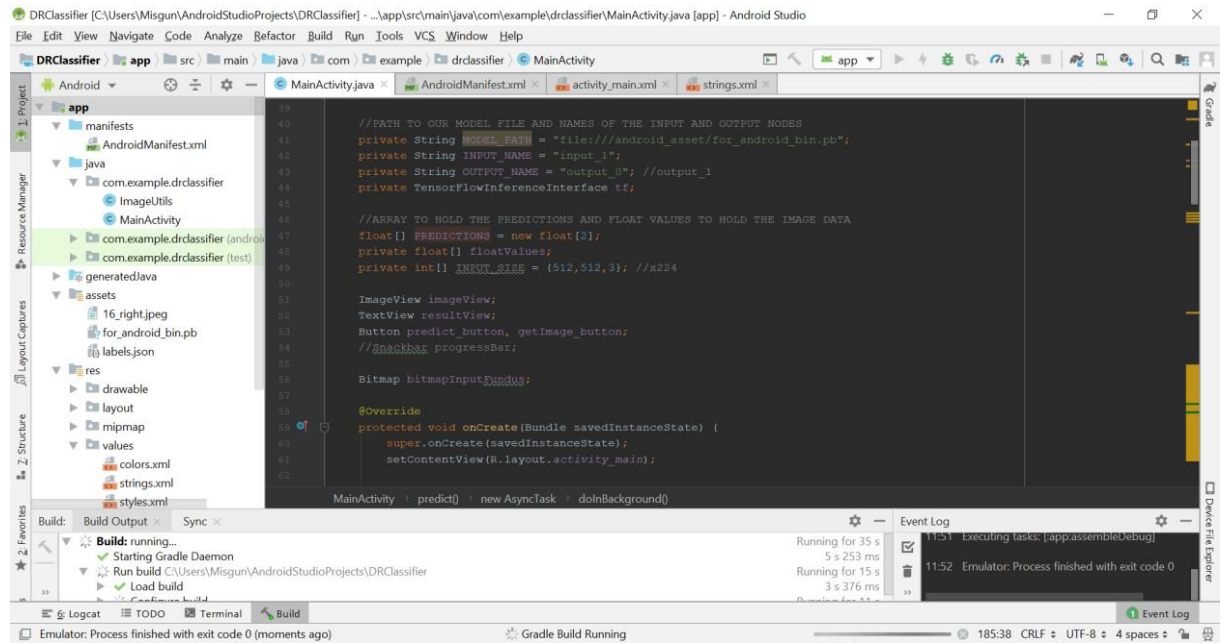

Figure 2.5. Android studio development screen

## 2.5. CONCLUSION

In this chapter the concept of deep learning have been discussed by giving examples of deep learning architectures such as convolutional neural networks, recurrent neural networks. Training methods, including types of learning, loss functions and optimization algorithms, that have been used in various literatures works have been defined and discussed. The deep learning development environments that are used in data preparation, preprocessing, model designing, development and deployment are also discussed.

A convolutional neural network as is presented in this chapter can be used either in end-to-end learning or transfer learning with one the training methods discussed to detect or classify diabetic retinopathy from fundus images. Spyder can be used to design and train a convolutional neural network model, and Android studio can be used to deploy the trained convolutional neural network into a smartphone.

# 3. DATASETS AND PERFORMANCE METRICS

In this chapter, different fundus images dataset that have been used for diabetic retinopathy detection and classification will be discussed, and their contents and dataset size will be compared. The EyePacs fundus images dataset which is hosted in website of Kaggle [30] will be presented. And the various performance metrics that have been used in the diabetic retinopathy identification, detection and classification literature are defined and discussed thoroughly.

## 3.1. RETINAL FUNDUS IMAGE DATASETS DESCRIPTION

Salz et al [6] presented different diabetic retinopathy imaging techniques. Diabetic retinopathy or fundus images can be diagnosed by using different imaging techniques such as fundus photography, optical coherence tomography, and fluorescein angiography.

In the research and development of automated diagnosis of diabetic retinopathy many datasets have been collected and publicly shared for research purposes. These datasets differ in the number of classes, number of fundus images collected and quality of the images.

In a study to prevent diagnosable blindness, Foster and Resnikoff [8] put four strategies to fight the challenges the diagnoses process of diabetic retinopathy faces in order to implement treatment for preventable blindness; (1) Creating academic, public and governmental awareness of the effects of blindness and visual loss, and the fact that 75% diseases that cause blindness are preventable; (2) Automating and mobilizing existing techniques and methods; (3) Implementing district-specific and country-specific prioritizing strategies of diagnosing and treatment resources for a productive process; (4) Providing comprehensive, maintainable and fair diagnosis services of visual diseases at district level, which includes stuff training, distributing diagnosis and treatment resources, and infrastructure, such as health care centers, building. The collection of annotated datasets of fundus images has huge impact in creating awareness in the general public, as is done in the Kaggle [30] dataset by hosting a classification competition.

In Table 3.1, an overview of the public domain benchmark datasets: MESSIDOR, e-ophtha, Kaggle, DRIVE, STARE, DIARETDB1, CHASE, DRiDB, ORIGA, SCES, AREDS, REVIEW, EyePACS-1, RIM-ONE, DRISHTI-GS, ARIA, DRION-DB and SEED-DB is presented.

Table 3.1 Dataset for Diabetic Retinopathy Detection

| Dataset | Number of images | Resolution | Format | Annotation | Tasks |
|---|---|---|---|---|---|
| MESSIDOR [31] | 1200 | 1440x960, 2240x1488, 2304x1536 | TIFF | Image level | - DR grading<br>- Risk of DME |
| e-ophtha[32] | 148 (MAs),<br>233 (Normal non-MA)<br>47 (EXs),<br>35 (Normal non-EX) | - | - | Pixel level | -MAs detection<br>-EXs detection |
| Kaggle [30] | 80000 | - | | Image level (5 classes), No DR, Mild, Moderate, Severe, PDR | -DR grading |
| DRIVE [33] | 33 (Normal)<br>7 (Mild early DR stage) | 584x565 JPEG | - | Pixel level | -Vessels extraction |
| STARE [34] | 400 | 605x700 | - | - Pixel level | -13 retinal diseases<br>-Vessels extraction<br>-Optic nerve |
| DIARETDB1 [35, 36] | 5 (Normal)<br>84 (At least one NPDR sign) | 1500x1152 | - | Pixel level | -MAs |
| CHASE [37] | 28 | 1280x960 | - | Pixel level | - Vessels extraction |
| DRiDB [38] - | 50 | - | BMP | Pixel level | -MAs<br>-HMs<br>-HEs<br>-SEs<br>-Vessels extraction<br>-OD<br>-Macula |
| ORIGA[39] | 482 (Normal)<br>168 (Glaucomatous) | | - | Pixel level | -OD<br>-Optic cup<br>-Cup-to-Disc Ratio (CDR) |
| SCES [40] | 1630 (Normal) 46 (Glaucomatous) | | | Pixel level | -Cup-to-Disc Ratio (CDR) |
| AREDS [41] | 72000 | - | - | Image level | AMD stages |
| REVIEW [42] | 16 | - | - | Pixel level | -Vessels extraction |
| EyePACS-1 [43] | 10000 | | | - Image level | -Referable DR -MA |
| RIM-ONE [44] | 18 (Normal) 12 (Early glaucoma) 14 (Moderate glaucoma) 14 (Deep glaucoma) 11 (Ocular hypertension) | - | - | - Pixel level | -Optic nerve |
| DRISHTI-GS[45] | 31 (Normal) 70 (Glaucomatous) | 2896x1944 | - | - Pixel level | -OD segmentation -Optic cup(OC) segmentation |
| ARIA [46] | 16 (Normal) 92 (AMD) 59 (DR) | 768x576 | - | - Pixel level | -OD -Fovea location -Vessel extraction |
| DRIONS-DB[47] | 110 | 600x400 | - | - Pixel level | -OD |
| SEED-DB[48] | 192 (Normal) 43 (Glaucomatous) | 3504x2336 | - | - Pixel level | -OD -Optic cup(OC) |

### 3.2. EYE PACS FUNDUS IMAGES DATASET

Diabetic retinopathy can be diagnosed by using different imaging techniques such as fundus photography, optical coherence tomography, and fluorescein angiography [6]. The EyePacs fundus images dataset was collected from distributed clinics that deploy fundus photography cameras and send them using telemedicine technology.

EyePacs is a United States based eye screening program. It has over 600 organizations using EyePacs. It has a solution of telemedicine screening. Clinics using the service will be able to capture and upload retinal images and other patient information to the EyePacs online database. At the other end, experienced eye care practitioners interpret the uploaded information. And at last the diagnosis and recommendations are sent to the clinic in the form of a portable document format or other formats. After this process, patients will be able to receive appropriate referrals, follow-up and treatment from their respective clinics.

EyePacs provides a secure framework, evidence based process of validation, an automated retinal image quality assessment, and an adaptable technology. EyePacs follows an international accepted diabetic retinopathy grading guidelines. Using the automated retinal image quality assessment system, which is built in collaboration with Google Inc., it is able to automatically assess image quality using state-of-the-art Artificial Intelligence (AI) technology and annotate images. This system also notifies the clinic staff that are using EyePacs screening program, on how to collect better quality images in regards to cleaning smudges, adjusting the illumination, and improving focus.

In cooperation with the California Healthcare Foundation, EyePacs provided a diabetic retinopathy dataset for a Kaggle fundus images classification competition [30]. The dataset contains a total of 35126 fundus images and their corresponding label for model training. The fundus images were labeled using a consensus that was reached by Wilkinson et al [5]. The labels range from zero to four, where zero represents a healthy fundus image, one represents a mild diabetic retinopathy, two represents a moderate diabetic retinopathy, three represents a severe diabetic retinopathy, and four represents a proliferative diabetic retinopathy. The dataset also contains unlabeled fundus images for testing.

Due to the highly unbalanced nature of the Kaggle dataset [30], when applying machine learning or deep learning algorithms for diabetic retinopathy classification weight balancing or image augmentation is required. The dataset contains 25812 fundus images labelled zero for no apparent signs of retinopathy. There are 2442 fundus images labeled one for a mild diabetic retinopathy. It contains 5291 fundus images labelled two for moderate signs of retinopathy. There are 873 fundus images labelled as three for severe stage of retinopathy. And it contains 708 fundus images labelled four for prolific retinopathy.

### 3.3. PERFORMANCE METRICS

Performance of classification algorithms needs to be tested before using in real-world application areas. This prevents users or customers from using an automated system that may be predicting, classifying or computing the wrong class. In medical image classification problems performance of classifiers should be tested attentively because their results or predictions of deployed could be a matter of life and death.

In this section, the performance metrics that are usually used to assess diabetic retinopathy automated classification are defined. The common metrics for measurement of classification algorithms are: sensitivity (recall), specificity, accuracy, precision, F-Score, Area under the ROC curve and logloss. Performance of segmentation algorithms can be assessed using Intersection Over Union (**IOU**), overlapping error, boundary based evaluation and dice similarity coefficient.

Accuracy is defined as the ratio of the correctly classified values to the incorrectly classified instances. Its equation is given in Equation (3.1).

$$\text{Accuracy} = \frac{TP + TN}{TP + TN + FP + FN} \quad (3.1)$$

The accuracy performance measure only calculates the percentage of correctly classified instances. It doesn't compute its performance in relation to the expected measure. Kappa statistic calculates the accuracy of prediction taking into account the expected accuracy measure of a model. As can be in Equation (3.2) the kappa score is useful to compare performances of different classifiers too.

$$\text{Kappa} = \frac{Observed\ Accuracy - Expected\ Accuracy}{1 - Expected\ Accuracy} \quad (3.2)$$

True Positive (TP) is the number of positive inputs (e.g. having diabetic retinopathy) in a given dataset that are classified as having diabetic retinopathy, whereas True Negative or TN is the number of healthy or negative inputs in the given dataset that are classified as healthy. False Positive or FP and False Negative or FN stand for the number of positive and negative instances, respectively, which are wrongly classified. In case of diabetic retinopathy classification, an input or instance is either fundus image, patch or a pixel of fundus image depending on the task. Sensitivity, Equation (3.3), or SN or true positive rate or recall measures the fraction of correctly classified positive inputs, and specificity, Equation (3.4), or SP or true negative rate measures the fraction of correctly classified negative inputs and precision or positive predictive value measures the fraction of positive instances that are correctly classified.

$$\text{Sensitivity (or Recall)} = \frac{TP}{TP + FN} \quad (3.3)$$

$$\text{Specificity} = \frac{TN}{TN + FP} \tag{3.4}$$

$$\text{Precision} = \frac{TP}{TP + FP} \tag{3.5}$$

$$\text{F Score} = \frac{2 \times Precision \times Recall}{Precision + Recall} \tag{3.6}$$

Equation (3.5) defines precision. F-score (or F) combines precision and recall as can be seen in Equation (3.6).

Receiver Operating Characteristic Curve (ROC) shows the plot of sensitivity against the false positive rate. It shows the relationship between sensitivity and specificity. Area under ROC curve (AUC), which is commonly used as an important metric of performance, always results in the values between 0 and 1; the closer the AUC to 1, the better the performance. Logarithmic loss (log loss) determines the accuracy of a classifier by penalizing false classifications. To calculate the log loss, the classifier is first designed to output the probability of each class instead just outputting the classified class. Log loss is given in Equation (3.7), where n is the size of dataset, m is the number of classes, $y_{ij}$ takes a binary result for whether or not class j is the correct class for input i, and $p_{ij}$ is the probability of assigning label j to input i.

$$\text{Logloss} = -\sum_{i=1}^{n} \sum_{j=1}^{m} y_{ij} \log p_{ij} \tag{3.7}$$

### 3.4. CONCLUSION

In this chapter different fundus images dataset that have been heavily used for diabetic retinopathy detection and classification works are discussed and their contents have been compared. The EyePacs fundus images dataset which is hosted in website of Kaggle [30] is presented. And performance measure of diabetic retinopathy detection and classification such as accuracy, sensitivity and specificity that have been deployed in the literature are discussed thoroughly.

# 4. LITERATURE REVIEW

In this chapter, automatic identification, detection and classification of diabetic retinopathy works will be discussed. Medical image classification can be done in two ways. It can be implemented using feature segmentation, detection and classification, or medical images can be classified using image level classification, where the whole image is considered and studied for classification. Due to the shortage of ophthalmologists, and healthcare centers around the world, mobile diagnosis service of diabetic retinopathy have also been implemented and researched. Smartphone based camera assisted diabetic retinopathy detection and classification will also be presented in this chapter.

Diabetic retinopathy lesions such as macula edema, exudates, microaneurysms, and hemorrhages have been used for feature based detection, and classification. In image level classification of diabetic retinopathy, deep learning models such as convolutional neural networks and deep belief neural networks have been used. Convolutional neural networks can be trained and used from scratch, which employs end-to-end training, or transfer learning can be employed for faster training and reduced training dataset. Transfer learning is the applicability of already trained neural network models in classifying previously unseen dataset. Transfer learning can be useful in medical image classification because healthcare sector of automated image classification suffers from annotated training data insufficiency. For clarity, Figure 4.1 presents a graphical representation of diabetic retinopathy detection techniques which accept retinal fundus images for image preprocessing, and detection purposes.

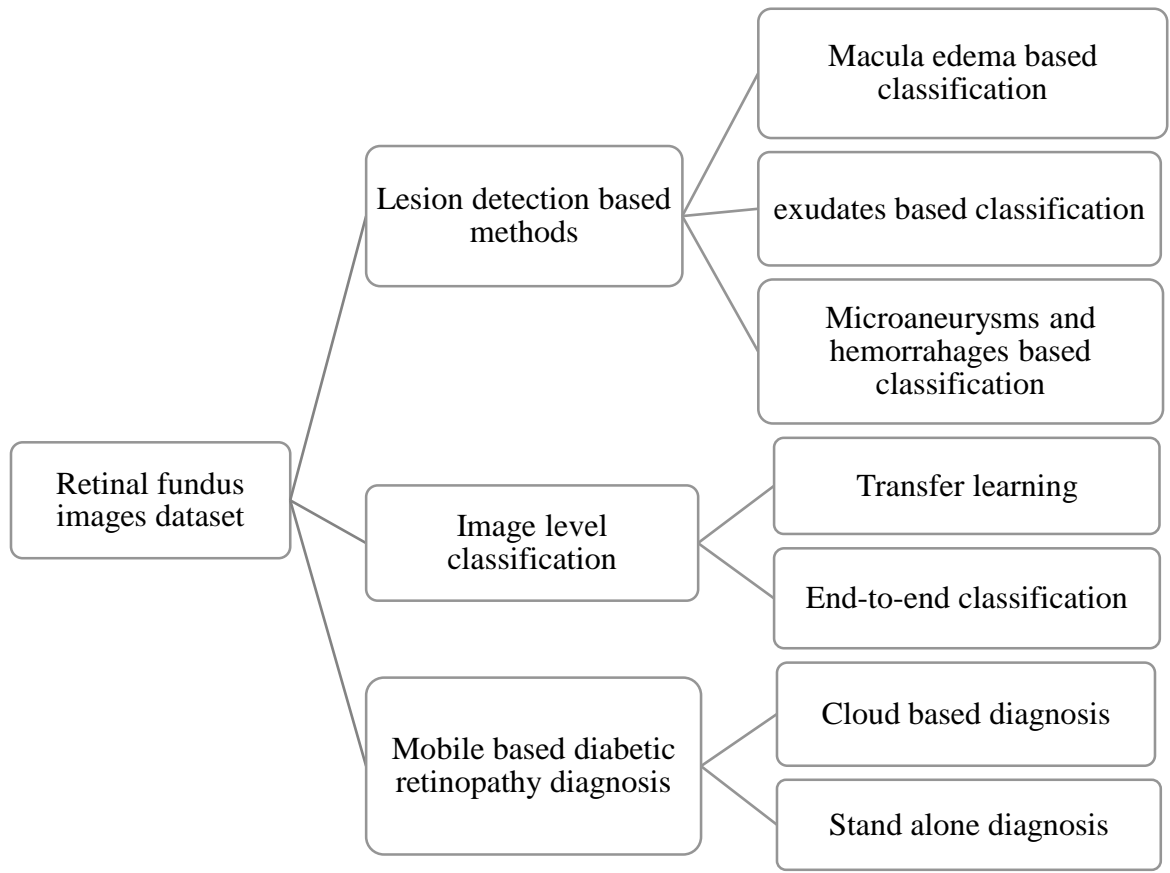

Figure 4.1 Literature works on automatic diabetic retinopathy diagnosis.

One of the main issues with incorporating deep learning in medical image analyses is the shortage of available labelled training dataset [11, 12]. Transfer learning techniques have gained wider acceptance because of the unavailability of labelled training data in the design and training of deep convolutional neural network models [13, 14]. In [15], annotated training data insufficiency was identified as the main challenge of applying deep learning models in the healthcare automation industry. Furthermore, Altaf et al. [15] recommended that methods, such as transfer learning, that exploit deep learning using reduced data need to be devised and implemented.

## 4.1. LESION DETECTION BASED CLASSIFICATION

Diabetic retinopathy detection can be automated using lesions extraction and classification. Macula edema, exudates, microaneurysms, and hemorrhages can be used as lesions for automated diabetic retinopathy detection. As can be seen in Figure 4.2, lesion detection based classification of diabetic retinopathy uses fundus images as input and HEF and machine learning for features extraction. In this section, lesion detection based automatic identification and classification methods of diabetic retinopathy are surveyed, presented, and compared.

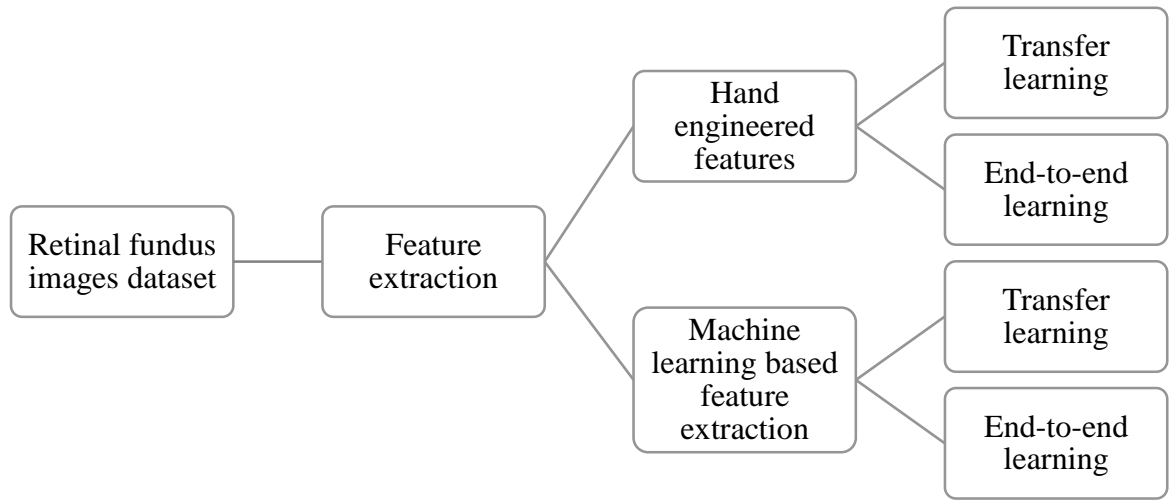

Figure 4.2 Categorization of lesion detection based diabetic retinopathy detection.

### 4.1.1. MACULA EDEMA BASED CLASSIFICATION

Macula is the central area and functional center of the retina where light is focused by cornea and lens structures in the front of the eye. The macula provides with the ability to read and see in great detail whereas the rest of the retina provides peripheral vision.

Diabetic Macula Edema (DME) is a diabetic retinopathy complication that occurs when the retinal capillaries become preamble and leakage occurs around macula; the macula swells and thickens when vessels fluid and blood gets into the retina.

Macula edema feature extraction based automatic recognition, identification and classification of diabetic retinopathy have been using convolutional neural networks, and Deep Belief Networks. Table 4.1 presents research works that have implemented diabetic retinopathy automated recognition using macula edema as a feature.

Table 4.1 Research works on diabetic retinopathy detection using macula edema as a feature

| Author(s) | Method | Training | Dataset name | Performance |
|---|---|---|---|---|
| Abràmoff et al. [49] | CNN inspired by AlexNet | End-to-end | MESSIDOR-2 | SN=100 |
| Mo et al. [50] |  | End-to-end | HEI-MED | SN=92.55, F=84.99 |

| | Cascaded FCRN | | e-ophtha | SN=92.27, F=90.53 |
|---|---|---|---|---|
| Perdomo et al. [51] | Patches based CNN model | Transfer learning | MESSIDOR | SN=56.5, SP=92.8 DME ACC=77 DME loss=0.78 |
| Burlina et al. [52] | CNN based on Over-Feat | Transfer learning | NIH AREDS | SN=Between 90.9-93.4 SP=Between 89.9-95.6 ACC=Between 92-95 |
| Al-Bander et al. [53] | CNN model with 3 Conv. Blocks and one FC block | End-to-end | MESSIDOR | SN=74.7, SP=95 |
| Ting et al. [54] | CNN | End-to-end | (35948 images) | SN=93.2, SP=88.7 AUC=0.931 |
| Arunkumar and Karthigaikumar [55] | DBN for training and multiclass SVM as classifier | End-to-end | ARIA | SN= 79.32, SP=97.89 ACC=96.73 |

### 4.1.2. EXUDATE BASED CLASSIFICATION

Exudate detection and extraction can be useful for diabetic retinopathy classification, but the process of exudate detection can be a challenging task because of significant variation in exudates size, shapes and contrast levels.

Convolutional neural networks have been used for the localization, segmentation and extraction of exudates for diabetic retinopathy detection and classification. Table 4.2 presents a comparative analysis of research works that have employed exudate identification techniques to automatically diagnose diabetic retinopathy.

Table 4.2 Research works on diabetic retinopathy detection using exudates as a feature

| Author(s) | Method | Training | Dataset | Performance |
|---|---|---|---|---|
| Prentasic and Loncaric [56] | 11-layers CNN, OD and Vessel | End-to-end | DRiDB | SN=78, F=78 |

|  | maps |  |  |  |
|---|---|---|---|---|
| Perdomo et al. [57] | Patches based LeNet CNN | Transfer learning | e-ophtha (40% of patches) | SN=99.8, SP=99.6 ACC=99.6 |
| Gondal et al. [58] | Octree based CNN model with CAM | Transfer learning | DIARETDB1 | SN:HE =100, SE=90.0 AUC=0.954 |
| Quellec et al. [59] | Octree based CNN(net A) | Transfer learning | DIARETDB1 | AUC:HE=0.735, SE=0.809 |
|  | Octree based CNN(net B) |  |  | AUC:HE =0.974, SE =0.963 |

### 4.1.3. MICROANEURYSMS AND HEMORRHAGES BASED CLASSIFICATION

Microaneurysms are seen in the starting periods of DR and retinal damage. Unusual escape of blood from retinal blood vessels around macula results in microaneurysms which are small in size. They appear sharp-edged as red spots. After time, greater size of microaneurysms having irregular shapes will be spotted. These are called haemorrhages or dot-and-blot because of their unusual margins. Haemorrhages are caused when walls of weak capillaries get broken.

Microaneurysms and haemorrhages have also been investigated using deep learning approaches as signs of diabetic retinopathy as is presented in this section. A comparative analysis of microaneurysms and haemorrhages based diabetic retinopathy identification and classification is given in Table 4.3. Convolutional neural networks and auto encoders have been used for microaneurysms and haemorrhages extraction based diabetic retinopathy classification

In terms of sensitivity, specificity, AUC and accuracy, the CNN based technique by Haloi [60] seems to outperform other methods for MA detection due using pixel augmentation instead of image based augmentation. Among CNN based methods, the methods by Gondal et al. [58] and Quellec et al. [59] are computationally efficient and jointly detect referable DR and red features.

Table 4.3 Research works on diabetic retinopathy detection using microaneurysms and haemorrhages as features

| Author(s) | Method | Training | Dataset | Performance |
|---|---|---|---|---|
| Haloi [60] | 9-layers CNN | End-to-end | MESSIDOR | SN=97, SP=95 AUC=0.982 ACC=95.4 |
| | | | ROC | AUC=0.98 |
| van Grinsven et al. [61] | Patches based selective sampling | End-to-end | Kaggle [30] | SN=84.8, SP=90.4 AUC=0.917 |
| | | | MESSIDOR | SN=93.1, SP=91.5 AUC=0.979 |
| Gondal et al. [58] | Octree based CNN model | Transfer learning | DIARETDB1 | SN: -HM=91 -Red small dots=52 |
| Quellec et al. [59] | Octree based CNN (net B) | Transfer learning | DIARETDB1 | AUC: -HM=0.999 -Red small dots=0.912 -Red small dots +HM=0.97 |
| Orlando et al. [62] | HEF + CNN features and RF classifier | End-to-end | DIARETDB1 | CPM=0.4884 |
| | | | MESSIDOR | SN=91.09, SP=50 AUC=0.8932 |
| Shan and Li [63] | Patches based SSAE | Transfer learning | DIARETDB | SP=91.6 F=91.3 ACC=91.38 |

### 4.2. IMAGE LEVEL CLASSIFICATION

Diabetic retinopathy can be diagnosed by employing image level image recognition, detection and identification techniques on fundus images dataset. Convolutional neural networks have been used to train neural models on training dataset. The importance of this techniques to detect diabetic retinopathy for referral and treatment purposes. If a fundus image is classified as having diabetic retinopathy the corresponding patient's eye can treated so that vision loss or impairment can be prevented.

After collecting 13673 fundus images, from 23 provinces in China, Li et al. [64] annotated the images into six classes, namely no diabetic retinopathy, mild, moderate, severe, proliferative, and ungradable. Li et al. [64] made the annotated dataset publicly available, and can be freely downloaded [65].

Nagasawa et al. [66] used deep convolutional neural networks to classify fundus images into healthy and referable untreated proliferative DR. A Sensitivity (SN) of 94.7%, Specificity (SP) of 97.2% and 0.969 Area Under the ROC Curve (AUC) was reached on 378 images collected using ultra-wide field scanning laser ophthalmoscope and fluorescein angiography (FA).

Inspired by real-life diagnosis of DR by ophthalmologists, Zeng et al. [67] developed transfer learning based Siamese-like Convolutional Neural Networks (CNNS) that accepts left and right eyes' fundus images at the same time, for binary and multi class classification of fundus images collected from the Kaggle dataset [30]. Zeng et al. [67] reached at a kappa score of 0.829 on a five stage classification.

Zhao et al. [68] used Residual Neural Network (ResNet) for feature extraction, attention mechanism to focus on important features, and bilinear net for classification with a weighted loss function to classify the Kaggle dataset [30] into five classes.

Verbraak et al. [69] showed that a United States (US) Food and Drug Administration (FDA) approved Artificial Intelligence (AI) based DR detection device [70] scored 100% SN in detecting on 16 fundus images diagnosed with vision threatening DR (VTDR). The device's performance was also compared with fundus images collected from 1425 type-2 diabetes patients graded according to the International Clinical Diabetic Retinopathy Severity Scale (ICDR) [5].

Performance of a commercially available deep learning based referable DR and referable Age-Related Macular Degeneration (AMD) recogniser system was validated in [71]. After the system reached AUC of 97.5%, SE of 92.0% and SP of 92.1% in detecting referable DR on 1200 fundus images, and AUC of 92.7%, SE of 85.8%, and SP of 86.0% in referable AMD recognition from 133821 fundus images, Gonzalez-Gonzalo et al. [71] concluded that although the system ignored intermediary disease stages, deep learning can help to introduce faster eye diseases screening, and accuracy can be improved with a computer assisted human diagnosis deployment.

Hagos and Kant 2019 [72] employed transfer learning on a pre-trained inception-V3 convolutional neural network to combat the challenge of annotated training data insufficiency in applying deep learning to medical images classification. After testing on a previously unseen dataset of 5000 fundus images which was subsampled from the Kaggle [30] dataset, Hagos and Kant 2019 [72] reached at an accuracy score of 90.9%.

In Table 4.4, a comparative analysis of research works that have employed image level classification on fundus images to classify diabetic retinopathy stages is presented.

Table 4.4 Research works on diabetic retinopathy detection using image level classification

| Author(s) | Method | Training | Dataset | Performance |
| --- | --- | --- | --- | --- |
| Gulshan et al. [73] | Inception-v3 CNN | Transfer learning | EyePACS-1 | SN=90.3% SP= 90% AUC0.991 |
| Colas et al. [74] | CNN model | End-to-end | Kaggle [30] | SN=96.2% SP= 66.6 AUC= 0.946 |
| Quellec et al. [59] | Ensemble net A, net B, AlexNet | Transfer learning | DIARETDB1 | AUC=0.954 |
| | | | Kaggle [30] | AUC=0.955 |
| | | | e-ophtha | AUC=0.949 |
| Costa and Campilho [75] | Sparse CNN | End-to-end | MESSIDOR(20% of images ) | AUC=0.90 |
| | | | DR1(20% of images ) | AUC=0.93 |
| | | | DR2(20% of images) | AUC=0.97 |
| Pratt et al. [76] | 13-layers CNN | End-to-end | Kaggle [30] | SN=95% ACC=75% |
| Gargeya and Leng [77] | ResNet+Gradient boosting tree | End-to-end | MESSIDOR-2 | AUC=0.94 |
| | | | e-ophtha | AUC= 0.95 |
| Ting et al. [54] | CNN | End-to-end | (71896 images) | SN=90.5 SP= 91.6 AUC= 0.936 |

| Abràmoff et al. [49] | CNN | End-to-end | MESSIDOR-2 | SN=96.8 SP=87 AUC=0.980 |
|---|---|---|---|---|
| Mansour [78] | AlexNet/SVM as classifier | Transfer learning | Kaggle [30] | SN=100 SP=93 ACC=97.93 |
| Orlando et al. [62] | HEF + CNN features and RF classifier | End-to-end | MESSIDOR | SN=97.21 SP=50 AUC=0.9347 |
| Nagasawa et al. [66] | CNN | End-to-End | 378 images | SN= 94.7% SP= 97.2% AUC=0.969 |
| Zeng et al. [67] | CNN | End-to-End | Kaggle [30] | kappa Score= 0.829 |
| Zhao et al. [68] | ResNet for feature extraction, attention mechanism to focus on important features, and bilinear net for classification | End-to-End | Kaggle [30] | Average of Classification = 0.5431 |
| Verbraak et al. [69] | AI based diabetic detection device validation | End-to-End | 16 images | SN = 100% on vision threatening diabetic retinopathy |
| Gonzalez-Gonzalo et al. [71] | Performance validation of deep learning based diabetic detection device validation | End-to-End | 1200 fundus images | AUC = 97.5% SN = 92.0% SP = 92.1% |
| | | | 133821 fundus images | AUC = 92.7% SN = 85.8% SP = 86.0% |

| Hagos and Kant [72] | Pre-trained Inception-V3 network | Transfer learning | 2600 fundus images for training; and 5000 fundus images for validation | ACC = 90.9% |

### 4.3. PHONE BASED DIABETIC RETINOPATHY DIAGNOSIS

Diabetes patients that reside in rural and remote areas usually suffer from delayed diagnosis of DR because of the expensive deployment of diagnosing equipment, and shortage of ophthalmologists and health care centers. This leads to losing access to early treatment. Foster and Resnikoff [8] put four strategies to fight the challenges the diagnoses process of diabetic retinopathy faces in order to implement treatment for preventable blindness; (1) Creating academic, public and governmental awareness of the effects of blindness and visual loss, and the fact that 75% diseases that cause blindness are preventable; (2) Automating and mobilizing existing techniques and methods; (3) Implementing district-specific and country-specific prioritizing strategies of diagnosing and treatment resources for a productive process; (4) Providing comprehensive, maintainable and fair diagnosis services of visual diseases at district level, which includes stuff training, distributing diagnosis and treatment resources, and infrastructure, such as health care centers, building.

As is put by Foster and Resnikoff [8] to automate and mobilize existing techniques and methods, a smart phone device with mobile retinal image capturing cameras could be used to diagnose diabetic retinopathy in remote and rural areas. This can solve the cost and time challenges associated with traditional diagnosis of diabetic retinopathy.

One of the strategies suggested by Foster and Resnikoff [8] for prevention of treatable blindness was the distribution of maintainable and unbiased eye care services at district level. Tele-retina has been implemented to capture retinal images by non-experts from remote areas and transferring to ophthalmic centre for diabetic retinopathy diagnosis [79, 80, and 81]. Tele retina needs a strong network connection between patients and health care centres. Teleophthalmology, which includes analysis of collected images, can be implemented to provide DR diagnosis remotely [82].

One of the main challenges of deploying diabetic retinopathy automatic diagnosis on a smartphone is the quality of the retinal images captured and the limited processing power and memory inside. A speedy, easy to use and cheap retinal imaging device could be used as an add-on with a smartphone to collect fundus images of patients as in [9]. As can be seen in Figure 4.3, Kim et al. [9] proposed the CellScope retina with a custom software to be installed on a smart phone. There are five stages until the production of a wide-field retinal image: (1) A user by holding the device in hand points it at the retina of a patient. (2) In built light emitting diode is used to illuminate the retina. (3) A green dot is used as a fixation target. (4) Each individual image is approximately $50^0$. With the current automated arrangement, five overlapping images are captured of the central, superior, nasal, temporal, and inferior retina. (5) A custom software is used to merge the collected images into one and create a wide-field image of the retina spanning approximately $100^0$.

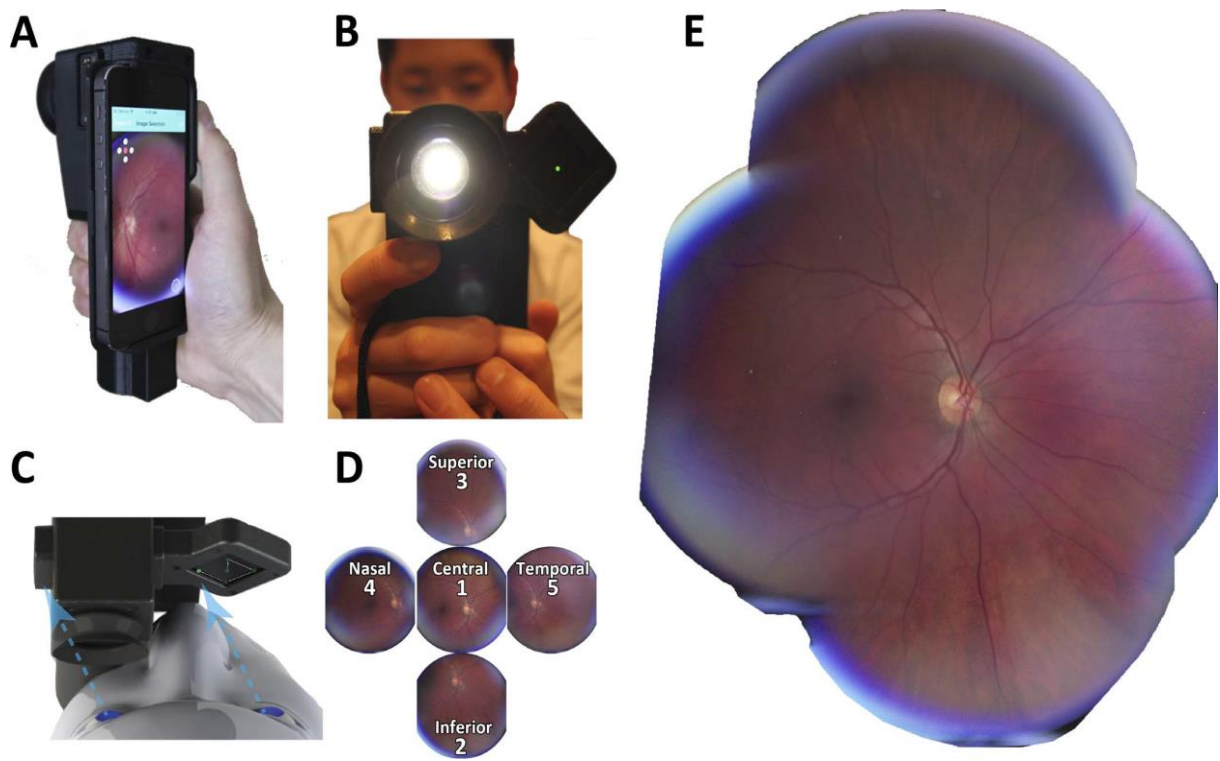

Figure 4.3 CellScope retina: (A, B) device held in hand. (B) In built light emitting diode is used to illuminate the retina. (C) A green dot is used as a fixation target. (D) Each individual image is approximately $50^0$. With the current automated arrangement, five overlapping images are captured of the central, superior, nasal, temporal, and inferior retina. (E) A custom software is used to merge the collected images into one and create a wide-field image of the retina spanning approximately $100^0$. [9]

Smartphone based automatic diagnosis of diabetic retinopathy can be implemented in two ways. The first and easy way is an internet based diagnosis; and the second one is a stand-alone independent smartphone application software based diagnosis. In an internet based diagnosis, smartphone is only used to capture a fundus image of a patient with the help of an add-on retinal camera. After a fundus image is captured, it is sent to a professional ophthalmologist for diagnosis, and results will be sent be over the internet. This would require the user to have a stable internet connection for sending the fundus images, and receiving back the results of diagnosis. The challenge of low internet coverage can be solved using an independent smartphone application that is able to process a captured fundus image and automatically diagnose the stage of diabetic retinopathy without any network connection. A stand-alone smartphone application software could enable patients living in rural and remote areas to get automatic diagnosis of diabetic retinopathy easily. A smartphone based diagnosis of diabetic retinopathy generally follows the process put in Figure 4.4.

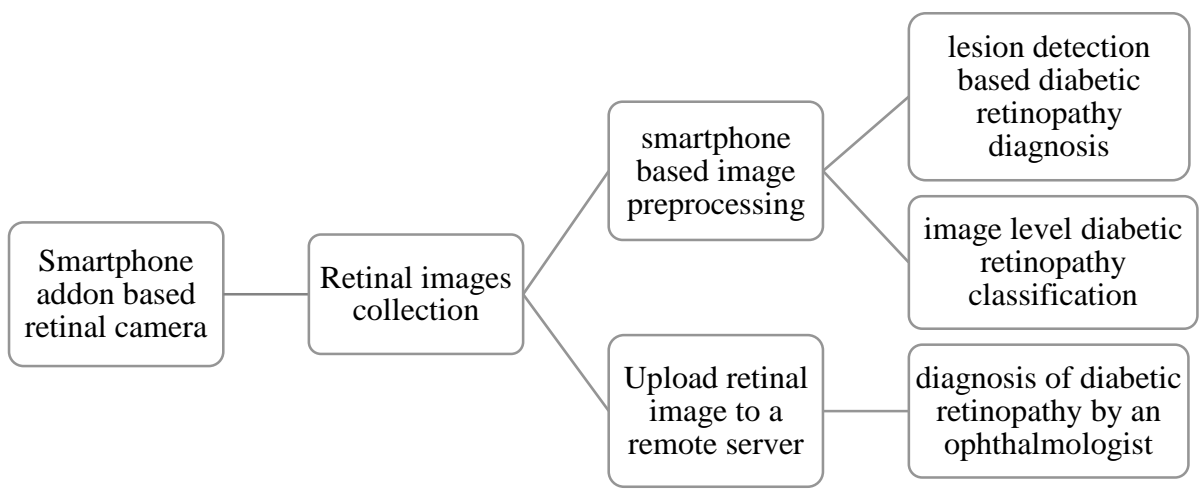

Figure 4.4 Process flow of a smartphone based diagnosis of diabetic retinopathy.

Prasanna et al. [83] proposed to develop a smartphone based DR detection system for remote areas. Fundus images were captured by attaching a handheld ophthalmoscope to a phone's camera; feature based classification was used to differentiate DR from healthy images; and it was also proposed to use cloud computing after mobile processing power showed limitation [83].

Xu et al. [84] implemented a vessel segmentation method on fundus images that can be collected using the help of a commercially available specialized cameras or downloaded from a publicly available datasets, and hosted it in an independent android smartphone system. An accuracy of 93.3% was reached on a validation dataset of 40 colour fundus images [84]. Figure 4.5 shows the proposed method in Xu et al. [84].

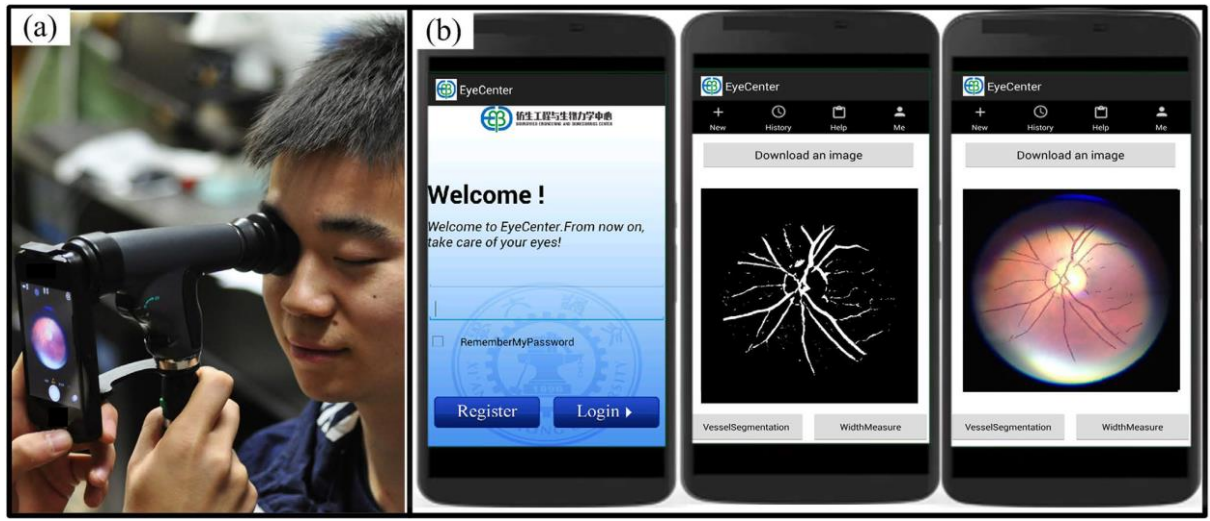

Figure 4.5 Smartphone app for fundus images' segmentation [84]

Rajalakshmi et al. [85] employed smartphone cameras to collect fundus images of diabetic participants; and diagnosed the images with the help of AI on a cloud-based diagnostic service. The smartphone and cloud-service joint automatic diagnosis of DR resulted in 95.8% SN and 80.2% SP for detecting any DR that includes moderate DR and Clinically Significant Macular Edema (CSME), and 99.1% SN and 80.4% SP in detecting VTDR, which includes severe and proliferative stages of DR.

Jamil et al. [86] validated the diagnostic readiness of fundus images collected using a Samsung Galaxy N9000 with 20 dioptre condensing lens by comparing their classification accuracies with that of slit lamp examination as the standard. Diagnostics on the fundus images collected with the smartphone and the slit lamp examination were conducted by two independent specialists [86].

In [87, 88], a combination of a smartphone, a Light Emitting Diode (LED), and a condensing lens were used for retinal images collection. Kashyap et al [87] proposed and implemented a Discrete Wavelet Transform (DWT) based feature extraction with Artificial Neural Networks (ANN) to identify stages of DR with a precision and recall of 63% and 57%, respectively on top 5 images. Kashyap et al. [88] employed histogram comparison of query retinal images with a database of healthy and affected eyes to detect DR with 62% precision and 53% recall. Table 4.5 summarises previous phone based works on diagnosis of diabetic retinopathy.

Table 4.5 Literature review of smartphone based diabetic retinopathy diagnosis and segmentation

| Author | Performed Operation | Deployment Environment | Camera | Validation dataset Size | Performance |
|---|---|---|---|---|---|
| Prasanna et al. [83] | Feature extraction based classification | Smartphone | ophthalmoscope | Not specified | Not specified |
| Xu et al. [84] | Vessel segmentation | Android Phone | Not specified | 40 | Acc=93.3% |
| Kashyap et al. [87] | Discrete Wavelet transform based artificial intelligence classification | Smartphone | Light emitting diode with condensing lens and phone camera | Not specified | precision = 63% recall = 57% |
| Rajalakshmi et al. [85] | Artificial intelligence and cloud based diagnostic service | Smartphone and cloud service diagnostics | Smartphone camera | Not specified | Sensitivity = 95.8%, Specificity = 80.2% for detecting any DR. Sensitivity = 99.1%, Specificity = 80.4% in detecting VTDR |

| Jamil et al. [86] | Validate accuracy of smartphone diagnosis | Samsung Galaxy N9000 | Samsung Galaxy N9000 camera with 20 dioptre condensing lens | Not specified | Not specified |
| Kashyap et al. [88] | Histogram comparison | Smartphone | Light emitting diode with condensing lens and phone camera | Not specified | precision = 62% recall = 53% |

## 4.4. CONCLUSIONS AND DISCUSSION

In this chapter, automatic identification, detection and classification of diabetic retinopathy works have been discussed. Diabetic retinopathy classification using fundus images can be performed in two separate techniques. It can be implemented using feature segmentation, detection and classification, or fundus images can be classified using image level classification, where the whole image is considered and studied for classification. Smartphone based camera assisted diabetic retinopathy detection and classification have also been presented in this chapter. A smartphone based automatic diagnosis of diabetic retinopathy solves the challenge of the shortage of trained ophthalmologists, and unavailability of enough medical equipment distribution around the world, specifically in rural and remote areas.

Diabetic retinopathy disease features such as macula edema, exudates, microaneurysms, and hemorrhages have been used for feature based detection, and classification. In image level classification of diabetic retinopathy, deep learning models such as convolutional neural networks and deep belief neural networks have been used. Convolutional neural networks can be trained and used from scratch, which employs end-to-end training, or transfer learning can be employed for faster training and reduced training dataset. Transfer learning is the applicability of already trained neural network models in classifying previously unseen dataset. Transfer learning can be useful in medical image classification because healthcare sector of automated image classification suffers from annotated training data insufficiency.

One of the main issues with incorporating deep learning in medical image analyses is the shortage of available labelled training dataset [11, 12]. Transfer learning techniques have gained wider acceptance because of the unavailability of labelled training data in the design and training of deep convolutional neural network models [13, 14]. In [15], annotated training data insufficiency was identified as the main challenge of applying deep learning models in the healthcare automation industry. Furthermore, Altaf et al. [15] recommended that methods, such as transfer learning, that exploit deep learning using reduced data need to be devised and implemented.

In this dissertation work, in order to avoid the challenge of the unavailability of enough annotated training data for the deployment of deep learning in medical image classification in general and in diabetic retinopathy classification specifically, three different approaches have been followed. And their results have been compared.

# 5. PROPOSED METHODOLOGY AND RESULTS

Diabetic retinopathy classification from fundus images can be done in two ways. It can be implemented using feature segmentation, detection and classification, or fundus images can be classified using image level classification, where the whole image is considered and studied for classification. In this thesis work, two different deep learning based models have been employed and compared in classifying diabetic retinopathy.

## 5.1. DATASET PREPARATION

The Kaggle diabetic retinopathy detection challenge dataset [30] contains colour fundus images that are labelled 0,1,2,3 or 4 for normal, mild, moderate, severe and prolific diabetic retinopathy, consecutively. Diabetic retinopathy progresses from mild, to moderate, severe non-proliferative diabetic retinopathy, to proliferative diabetic retinopathy. Two different subsets of the Kaggle dataset that consist of independent training and validation data were prepared for two different operations that are binary and multiclass classifications.

Subsamples of the EyePacs dataset that is uploaded on Kaggle diabetic retinopathy detection challenge [30] were used for model training and testing. 1000 fundus images, which consist of 200 fundus images selected from each of the five diabetic retinopathy stages, were used for model training. For model testing and validation, a total of 5000 fundus images subsample were selected representing each of the five stages of diabetic retinopathy. In order to save storage space the fundus images were resized to width of 200 pixels and height of 200 pixels.

The preprocessed fundus images were uploaded to Google Drive. The fundus images are distributed into training and validation directories as seen in Figure 5.1. The training directory contains five subdirectories for each of the five stages of diabetic retinopathy.

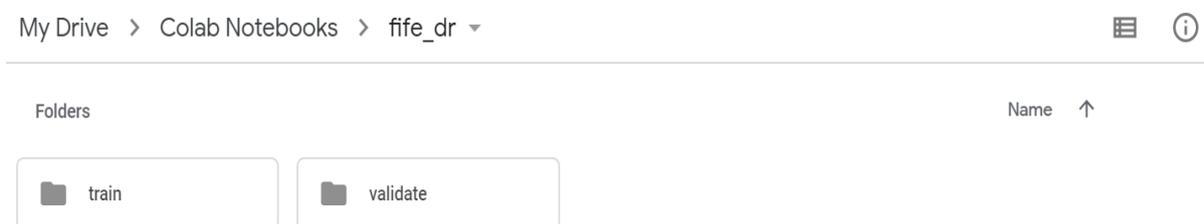

Figure 5.1. Fundus images distribution in Google Drive directory

As can be seen in Figure 5.2, all the fundus images have been put in their respective directories indicating labels. Each directory contains 200 fundus images, and the contents of a directory are labeled to the name of the directory. This is displayed in Figure 5.3. The same number of fundus images and folders are put in the validation folder too.

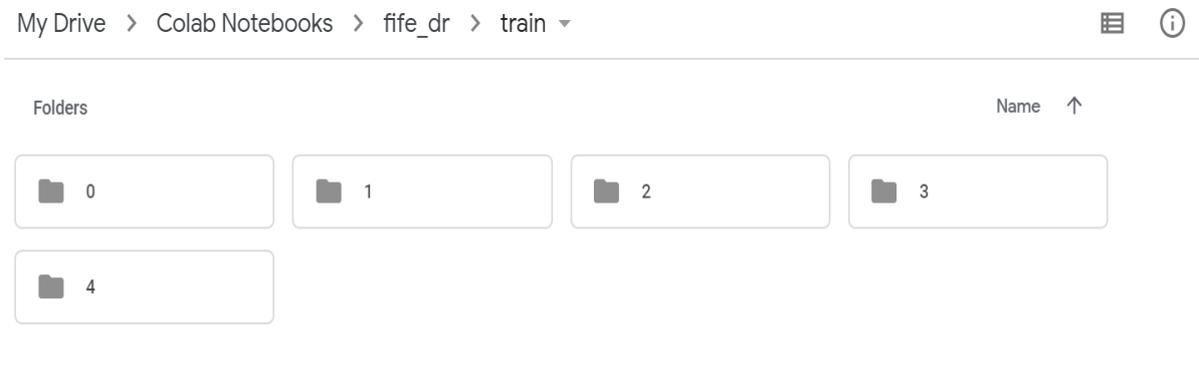

Figure 5.2. Distribution of fundus images inside training dataset for multiclass classification

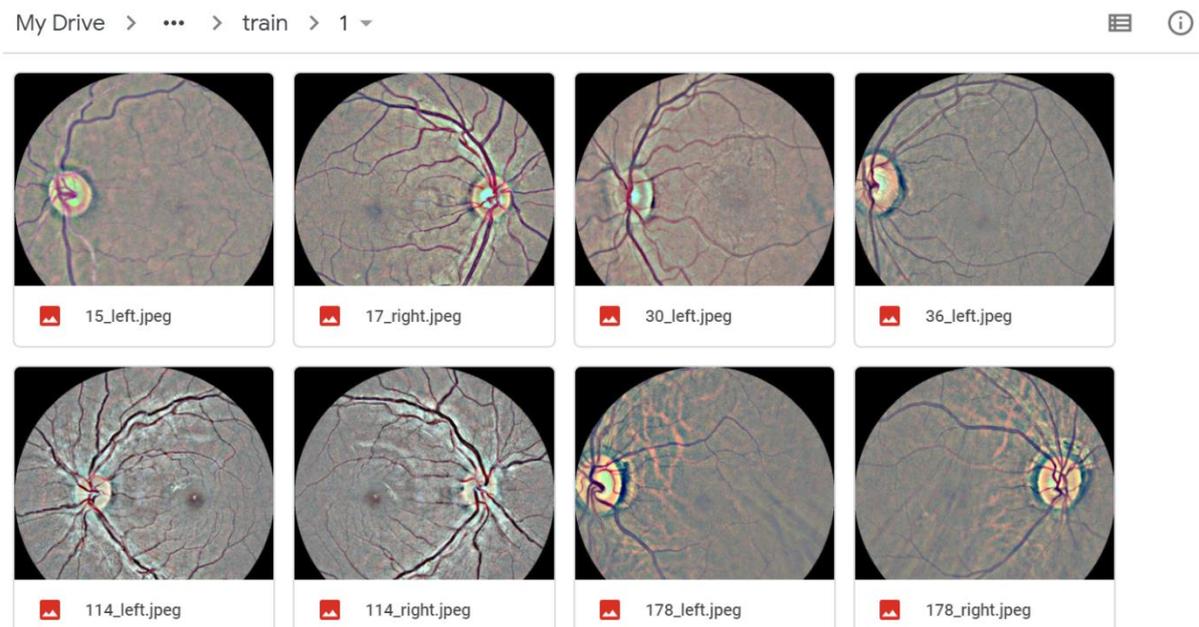

Figure 5.3. Fundus images labelled into stage one

## 5.2. IMAGE PRE-PROCESSING

After downloading the fundus images from the Kaggle dataset [30], the images may not be ready for implementing classification algorithm on them, or for deep learning training. Fundus images may contain noise that may obstruct a deep learning model from extracting the necessary features required for diabetic retinopathy classification. In order to avoid this challenge, image pre-processing techniques were employed on the selected fundus images.

Opencv-Python was used to remove black surrounding pixels from the input fundus images (See Appendix A.1). As can be seen in Appendix A.1, the local average was subtracted from each pixel [89]. After the local average was subtracted 90% of the cropped retinal image was cropped again, in order to avoid training with white rounding pixels. Result of the image pre-processing performed on a sample fundus image can be seen in Figure 5.4. Image pre-processing was implemented using Spyder IDE.

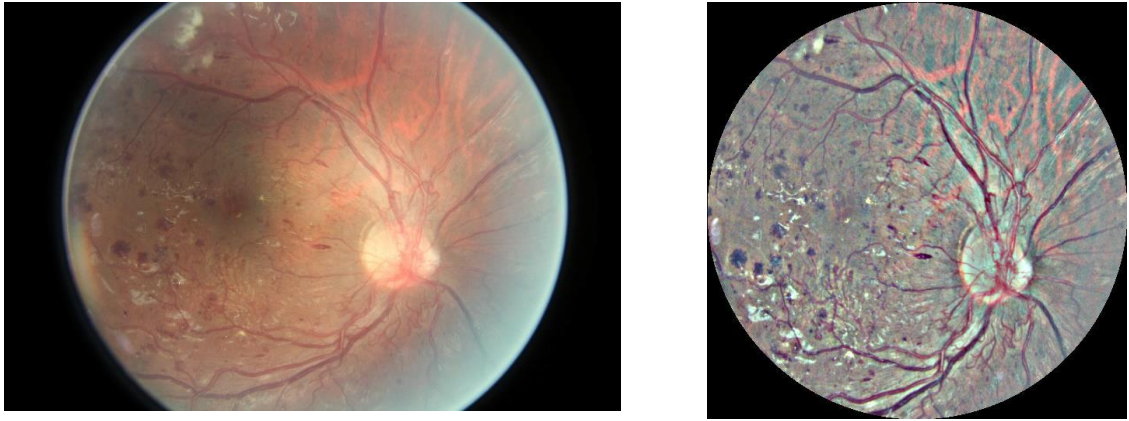

Figure 5.4. Input fundus image (left), pre-processed fundus image (right)

## 5.3. DEEP LEARNING MODELS

In this dissertation work, two different deep learning models have been designed and trained to classify diabetic retinopathy from a subsample of a fundus images dataset. Their architectures and performance scores have been reported in the next subsections.

### 5.3.1. TRANSFER LEARNING BASED MODEL

One of the main issues with incorporating deep learning in medical image analyses being the shortage of available labelled training dataset [11, 12], it has led to the incorporation of transfer learning in building deep convolutional neural network models [13, 14]. In [15] annotated training data insufficiency was identified as the main challenge of applying deep learning models in the healthcare automation industry. Furthermore, Altaf et al. [15] recommended that methods, such as transfer learning, that exploit deep learning using reduced training data need to be devised and implemented.

Transfer learning is the applicability of already trained neural network models in classifying previously unseen dataset. It includes importing a pre-trained deep learning model, extracting features of input dataset using the imported model and building a classifier on top of the extracted features. A convolutional neural network usually includes two main modules: a convolutional and a classifier module. When importing a pre-trained model, a user will be able to select which parts of the model to retrain or use as they are pre-trained.

## 5.3.1.1 MODEL ARCHITECTURE

In this dissertation work, a transfer learning based classification of diabetic retinopathy has been implemented on the inception-V3 deep neural network for binary classification. The classification task is performed using an ensemble one versus one binary classifiers among the five classes. For clarity, Figure 5.5 shows a clear process flow of how a transfer learning approach has been followed in order to classify diabetic retinopathy using input fundus images.

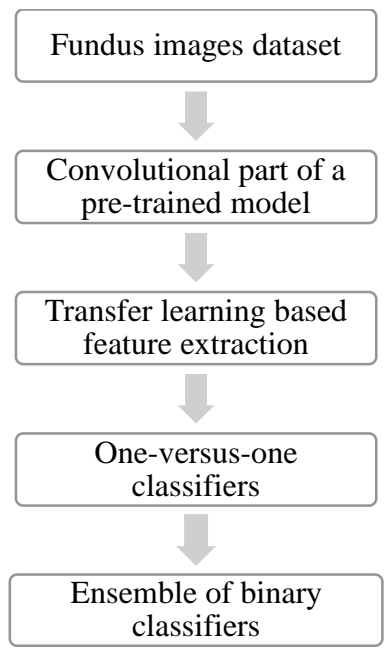

Figure 5.5. Implementation diagram of transfer learning based classifier

A pre-trained inception-V3 deep convolutional neural network has been used for transfer learning. It was trained on the ImageNet dataset. The whole of the convolutional part of the inception-V3 were on freeze so as not to retrain their parameters. Five independent binary one-versus-one classifiers were trained on top of the convolutional module. Each of the binary classifiers contained a fully connected rectified linear unit activated layer of 32 nodes was added as a classifier. In order to classify extracted features a Rectified Linear Unit (ReLu) activated layer was added to the end of the already trained model. Stochastic Gradient Descent (SGD) with an increasing learning rate of 0.0001 was implemented for training, and a Softmax activation was used at output nodes. Cosine loss function was used to calculate loss of the model because it has shown improved performance on datasets with limited size [90].

Results of five binary one-versus-one classifiers was combined to predict label of input image after training. The binary classifier models were to detect between normal and mild, normal and moderate, normal and severe, and between normal and proliferative. As proliferative is the advanced stage of diabetic retinopathy the final classifier would start identifying if an input is proliferative as can be seen in Figure 5.6.

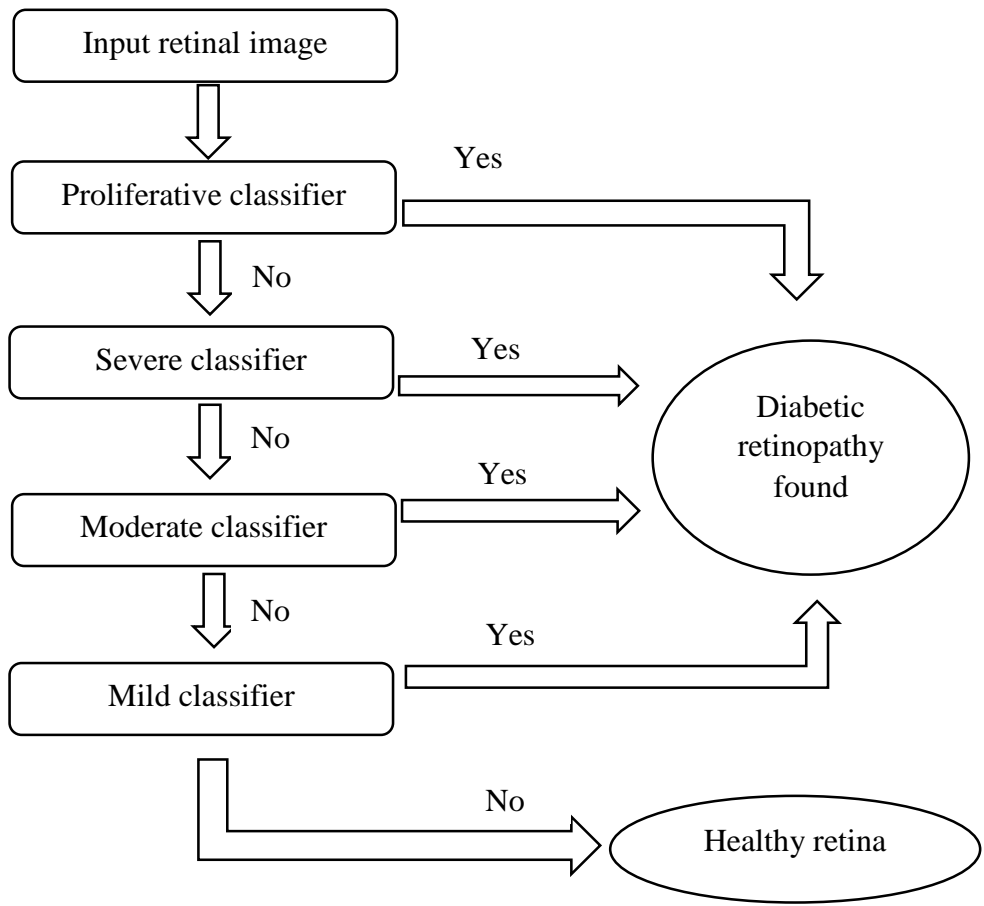

Figure 5.6. Overview of the combination of one-versus-one classifiers: the five binary classifiers are applied in sequence to classify a fundus image into proliferative, severe, moderate, mild or normal.

The inception-V3 was imported in order to take advantage of it design nature. It contains inception modules that are suitable to extract features of differing size in one stage of convolution. Inception modules as seen in Figure 5.7 are the unique features of an inception-v3 neural network.

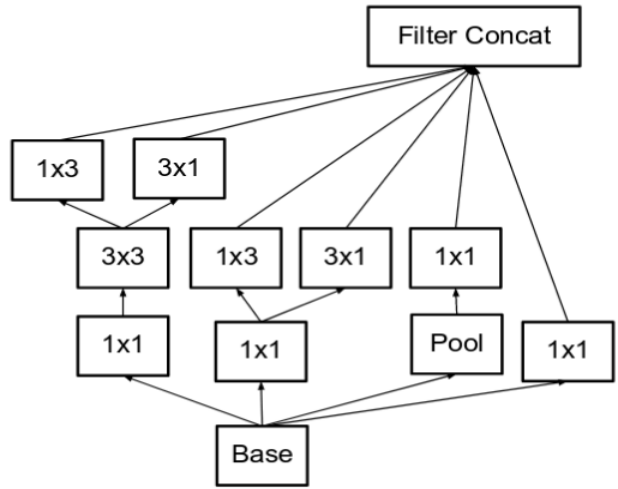

Figure 5.7. Inception module used in inception-V3

Szegedy et al. [22] designed the inception-V3 deep convolutional network by adding convolution factorization and auxiliary classifiers in inception modules. Number of network parameters was reduced by incorporating convolution factorization that replaces bigger convolutions with multiple smaller convolutions, and regularization effect was brought to the network by employing auxiliary classifiers. Inception networks are known for their use of inception modules. Inception modules are a collection of different sized filters working on the same level of convolution on an input image. Inception modules help to extract features that differ in size among images of the same class. The design nature of inception modules is helpful in diabetic retinopathy classification because features of diabetic retinopathy usually differ in size.

### 5.3.1.2 CLASSIFICATION RESULTS

The trained model's performance was tested on a previously unseen dataset of 5000 fundus images. The contents of the test dataset were 1000 previously unseen fundus images from each of the stage of diabetic retinopathy. Spyder with Python programming was used for model testing (See Appendix A.2). The classification result of the test dataset can be seen in Table 5.1.

Table 5.1. Classification result of ensemble of one-versus-one model

| Ground truth | Stage | Size of dataset | Predicted class | |
|---|---|---|---|---|
| | | | Healthy | Unhealthy |
| **Healthy** | 0 | 1000 | 966 | 34 |
| **Unhealthy** | 1 | 1000 | 51 | 949 |

|   | 2 | 1000 | 40 | 960 |
|---|---|------|----|----|
|   | 3 | 1000 | 42 | 958 |
|   | 4 | 1000 | 36 | 964 |

The sensitivity of the ensemble approach was 95.77%, and the specificity score was 96.60%. The accuracy of the proposed method became 95.94%.

In training each of the five models, at the end of each epoch, accuracy score of the resulting model was calculated on the test dataset of 5000 fundus images. A sample training history of the mild classifier can be seen in Figure 5.8.

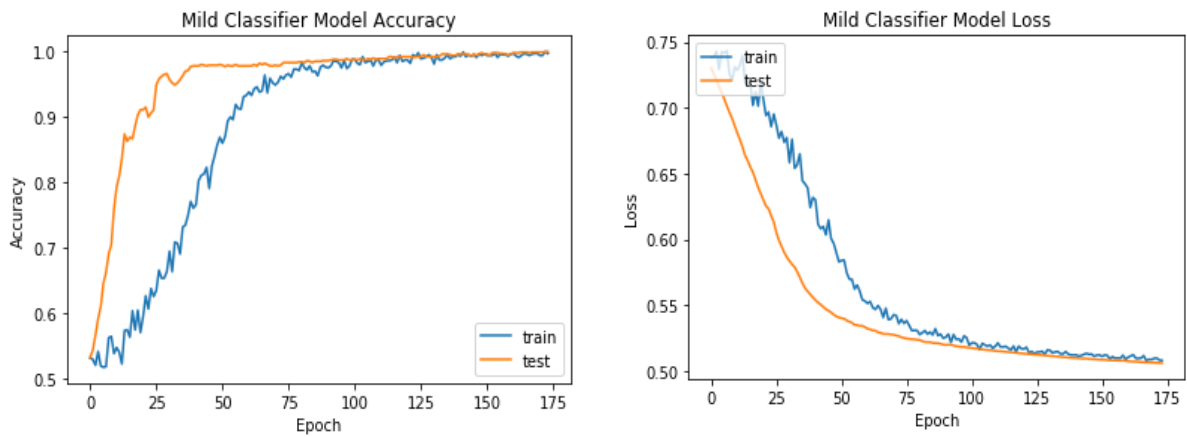

Figure 5.8. Mild classifier model training history: accuracy score (left), cosine loss (right)

In order to compare performance of two or more different models, the input and output settings such as the training dataset and the number and types of classes must be the same. Mohammadian et al. [91] and Hagos and Kant [72] have used transfer learning to classify fundus images in the Kaggle dataset [30] into healthy and unhealthy classes. A performance comparison of the proposed model with the models in [91] and [72] is presented in Table 5.2 depending on parameters of the model built such as training dataset size used, optimizer, learning rate, loss function, and whether data augmentation was employed or not. The proposed model has scored superior performance to on account of the selected model optimizer, which is SGD, and loss function, which is the cosine loss function, using a much reduced training data.

Table 5.2. Comparison of performance

| | Points of comparison | Mohammadian et al. [91] | Hagos and Kant [72] | Proposed method |
|---|---|---|---|---|
| Training parameters | Size of training data | More than 35126 fundus images | 2500 fundus images | 1000 fundus images |
| | Optimizer | ADAM | SGD | SGD |
| | Learning Rate | Not specified | Ascending rate of 0.0001 | Ascending rate of 0.0001 |
| | Loss function | Not specified | Cosine loss function | Cosine loss function |
| | Data Augmentation Used | Yes | No | No |
| Results | Accuracy | 87.12% | 90.6% | 95.94% |
| | Specificity | Not specified | Not specified | 96.60% |
| | Sensitivity | Not specified | Not specified | 95.77% |
| | Loss | Not specified | 3.94% | Not specified |

### 5.3.2. SMALL INCEPTION NETWORK

Inception modules have been observed to result in high performance in medical image classification [72]. For multiclass classification of diabetic retinopathy from fundus images, a new neural network that is based on inception modules has been designed and implemented in this dissertation network. The small inception network was combined with a one versus one binary classifiers. As seen in Equation (5.1), since there are five classes of diabetic retinopathy, after building a one-versus-one classifier ten classifiers were used where L is the total number of labels in the dataset.

$$\text{Number of Classifiers} = \frac{L * (L - 1)}{2} \quad (5.1)$$

#### 5.3.2.1 MODEL ARCHITECTURE

Two inception modules were used after three convolution operations were performed on input image. The overall architecture of the designed model is seen in Figure 5.9. Rectified Linear Unit (ReLu) was used as activation function across the whole small inception network. The inception module used can be seen in Figure 5.7. The input fundus images were resized to width of 200 and height of 200 to fit memory. The image preprocessing was performed using Spyder integrated development environment and the model was trained using Google Colaboratory [28].

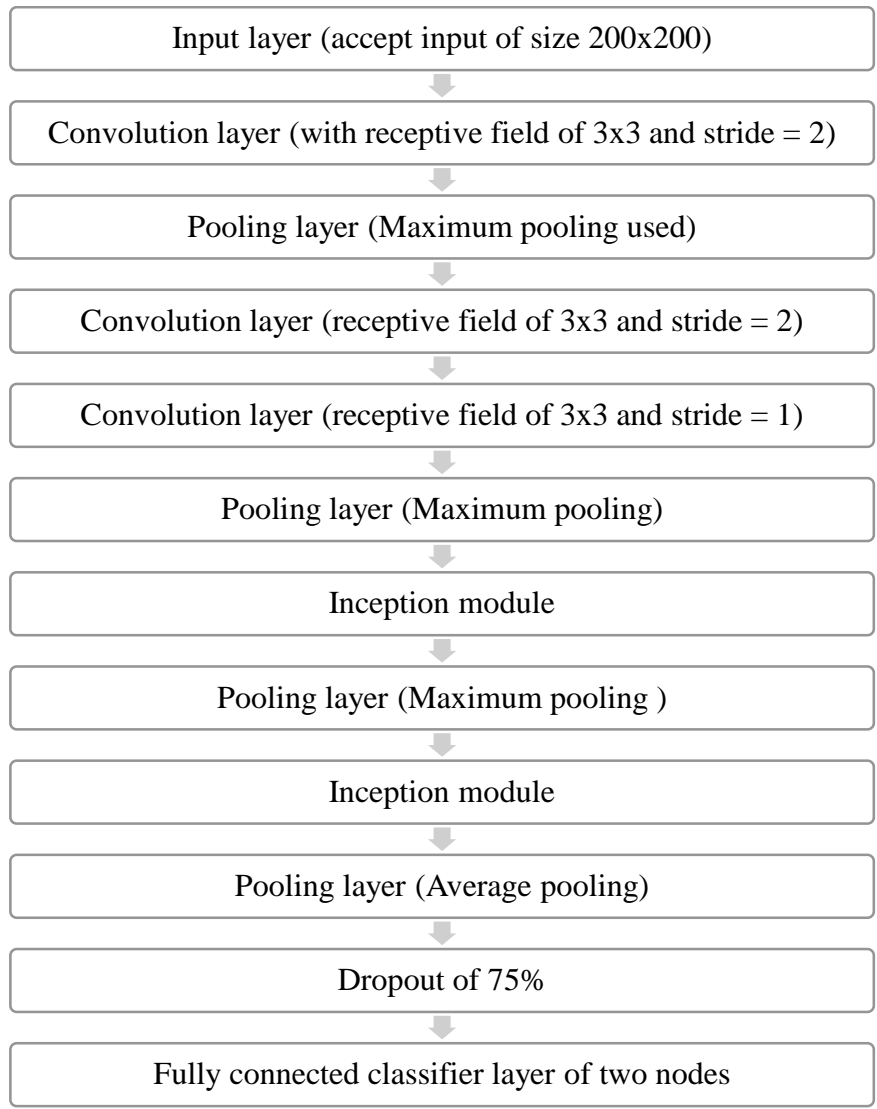

Figure 5.9. Model architecture of the small inception network

Ten binary classifiers classify on a one-versus-one classification basis as seen in Table 5.3. And predicted label will be computed using the results of all the ten binary classifiers (See appendix A.3)

The same small inception module architecture presented in Figure 5.7, with the same training parameter setting has been inserted across all the ten binary classifiers so their results can be easily compared, and aggregated to predict the final result of classification. Binary classifiers prediction probability needs to be aggregated by selecting the corresponding classifiers when calculating probability of a particular diabetic retinopathy stage. Classifiers zero, four, five and six would be used to calculate probability of a fundus image being diagnosed with a mild diabetic retinopathy.

Table 5.3 One-versus-one binary classifiers

| Classifier Number | Stage versus Stage |
|---|---|
| 0 | Normal versus mild |
| 1 | Normal versus moderate |
| 2 | Normal versus severe |
| 3 | Normal versus proliferative |
| 4 | mild versus moderate |
| 5 | mild versus severe |
| 6 | mild versus proliferative |
| 7 | moderate versus severe |
| 8 | moderate versus proliferative |
| 9 | severe versus proliferative |

In order to calculate probability of a fundus image of being diagnosed with one of the five diabetic retinopathy stages, we accumulate probability results of the corresponding four binary classifiers. Binary classifiers zero to three are used to compute the total probability of a single fundus image being a normal fundus image. Binary classifiers zero, four, five, and six are used to find the aggregate probability of a fundus image being diagnosed with a mild diabetic retinopathy. In order to compute the total probability of an image being labelled as moderate, binary classifiers one, four, seven and eight are used. Binary classifiers two, five, seven and nine are used to compute the total probability of an image's probability of being diagnosed with a severe diabetic retinopathy. In order to calculate the aggregate probability of a fundus image being diagnosed with a proliferative, binary classifiers three, six, eight and nine are used.

### 5.3.2.2 CLASSIFICATION RESULTS

Performance of the trained models was tested on a previously unseen dataset of 5000 fundus images which consists of 1000 images per diabetic retinopathy stage. The corresponding trained binary models were ensemble to predict classes of the five stage of diabetic retinopathy (See Appendix A.4). For clarity, classification results of the stages of diabetic retinopathy are presented in Table 5.4.

Table 5.4. Classification result of ensemble of one-versus-one model

| Ground truth | Size of input dataset | Predicted label | | | | |
|---|---|---|---|---|---|---|
| | | 0 | 1 | 2 | 3 | 4 |
| 0 | 1000 | 1000 | 0 | 0 | 0 | 0 |
| 1 | 1000 | 1 | 636 | 251 | 63 | 49 |
| 2 | 1000 | 0 | 133 | 670 | 104 | 93 |
| 3 | 1000 | 1 | 118 | 159 | 448 | 274 |
| 4 | 1000 | 0 | 111 | 190 | 135 | 564 |

Depending on the above classification result, the observed accuracy is 0.6636 using the performance metrics discussed in Chapter 3. And the kappa score is 0.612.

### 5.4. SMARTPHONE BASED DIAGNOSIS

Incorporating automatic diagnosis and classification of diabetic retinopathy would greatly help those that have less coverage of medical services because of the current wider acceptance of smartphones around the world. Smartphone based automatic diagnosis of diabetic retinopathy can be implemented in two ways. The first and easy way is an internet based diagnosis; and the second one is a stand-alone independent smartphone application software based diagnosis. In an internet based diagnosis, smartphone is only used to capture a fundus image of a patient with the help of an add-on retinal camera. After a fundus image is captured, it is sent to a professional ophthalmologist for diagnosis, and results will be sent be over the internet. This would require the user to have a stable internet connection for sending the fundus images, and receiving back the results of diagnosis. The challenge of low internet coverage can be solved using an independent smartphone application that is able to process a captured fundus image and automatically diagnose the stage of diabetic retinopathy without any network connection. A stand-alone smartphone application software could enable patients living in rural and remote areas to get automatic diagnosis of diabetic retinopathy easily.

#### 5.4.1. SYSTEM ARCHITECTURE

Two inputs are used in developing a smartphone based standalone diabetic retinopathy diagnosis. A trained deep learning model, and a retinal image captured with add on retinal camera are used in building the smartphone application. Figure 5.10 shows the process flow followed in developing the system.

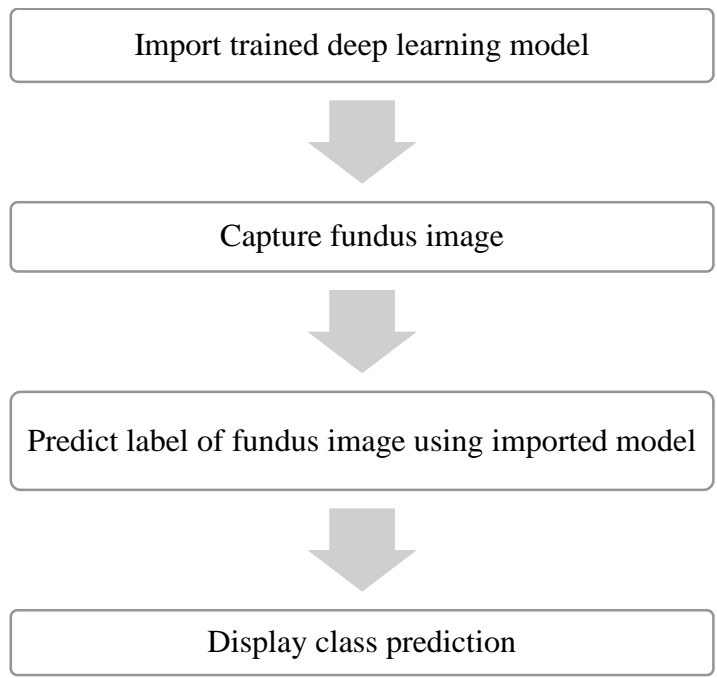

Figure 5.10. System architecture of a smartphone application for diabetic retinopathy diagnosis

The system was designed and built using Android Studio software. Java was used as a programming language. The user interface of the android based diabetic retinopathy diagnostic system is seen in Figure 5.11.

A retinal image is proposed to be captured using add on camera or from the smartphone's gallery by taping on the "GET" button. And the label of the imported retinal image is predicted by tapping on the "CLASSIFY" button. A typical result of prediction of a fundus image inputted from a smartphone's gallery is shown in Figure 5.12. The smartphone used was an Oppo realme-1 sub-brand, model CPH1859 with a ColorOS version 6.0 operating system, which is a fork of Google's Android operating system.

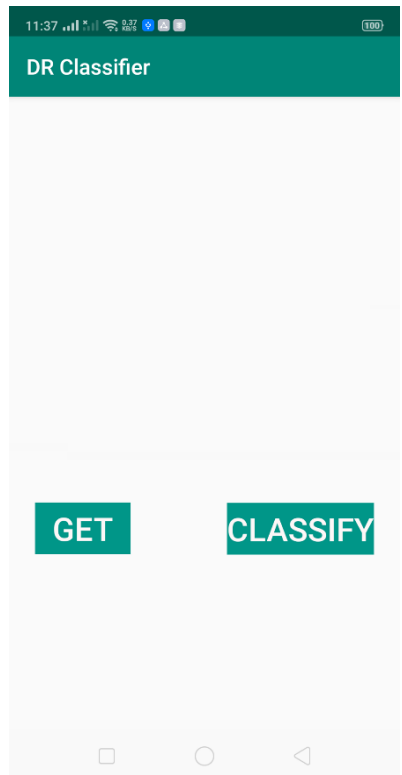

Figure 5.11 User interface of smartphone based diabetic retinopathy diagnosis. The "GET" button is used to import fundus image using add on camera or from the smartphone's gallery. The "CLASSIFY" button is used to predict the label of imported retinal image.

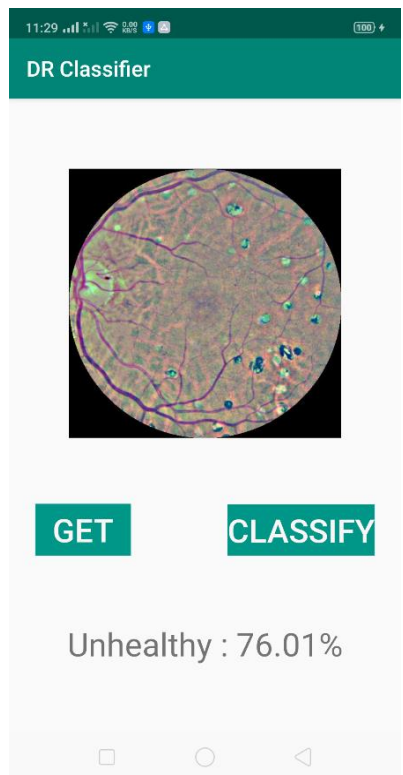

Figure 5.12. Prediction result of a smartphone based diabetic retinopathy diagnosis.

# 6. DISCUSSION AND CONCLUSION

In this dissertation work, image processing, deep learning techniques and android smartphone applications development have been combined in order to provide a standalone smartphone based automatic diagnosis of diabetic retinopathy. Diabetic persons living in remote and rural areas with less coverage of health center services and shortage of professional ophthalmologists only need an Android smartphone and a retinal camera add on for an instant DR diagnosis.

In the automatic diabetic retinopathy diagnosis literature various image processing and recognition techniques have been proposed, but they can be generally classified into two: feature extraction based and deep learning based. In feature extraction based classification of diabetic retinopathy lesions such as exudates, microaneurysms, hemorrhages and macular edema have been identified from retinal images. And in deep learning based detection works deep and narrow neural networks have been extensively used. A deep learning based approach was followed in this dissertation work, and two different models were designed and implemented in Python. Binary and multiclass classification were implemented for classifying fundus images. The binary classifier was implemented using transfer learning on a pre-trained deep learning model. A novel approach of ensemble one-versus-one based ten binary classifiers were used to perform multiclass classification. Both classifiers incorporated image processing techniques in order to exploit capabilities of deep learning in medical image classification from reduced training data. The trained binary classifier model was integrated with a standalone smartphone based diagnostic mobile application which was developed to give an easy and fast diagnoses of diabetic retinopathy in rural areas. The developed smartphone application doesn't need an internet connection or a trained professional in order to diagnose diabetic retinopathy. It is proposed that the fundus images captured by the smartphone are going to be collected using a special add on retinal camera.

A standalone smartphone based application can be used in order to provide DR early treatment and medication with a fast and easy-to-use automatic DR diagnostic service for diabetic persons living in rural and remote areas. In order to decrease the cost of deployment and periodic diagnosis, cheap add on retinal cameras that can be integrated with major smartphone operating systems need to be distributed.

# APPENDICES

## A. SOURCE CODE LISTINGS

### A.1 SOURCE CODE LISTINGS FOR IMAGE PREPROCESSING

```
"""
Created on Thu May 23 22:49:41 2019

By Misgina Tsighe Hagos
"""

import cv2
import numpy as np
from matplotlib import pyplot as plt
import os
from PIL import Image

def scaleRadius(img , scale):
    x=img[img.shape[0]//2,:,:].sum(1)
    r=(x>x.mean()/10).sum()/2
    s=scale*1.0/r
    return cv2.resize(img, ( 0 , 0 ) , fx=s , fy=s )

orig_address = "C:/Users/Misgun/Documents/Python Scripts/fully_graham_preped/4/"
output_address = "C:/Users/Misgun/Documents/Python Scripts/fully_graham_preped/4_preped/"
directory = os.fsencode(orig_address)

for im in os.listdir(directory):

    im = os.fsdecode(im)
    img_address = orig_address + im
    output_path = output_address + im
    colr_img = cv2.imread(img_address)
    colr_img = cv2.cvtColor(colr_img, cv2.COLOR_BGR2RGB)

    #Start Cropping here
    gray_img = cv2.cvtColor(colr_img, cv2.COLOR_BGR2GRAY)
    retval, thresh_gray = cv2.threshold(gray_img, thresh=50, maxval=255,
type=cv2.THRESH_BINARY)
    # Mask of non-black pixels (assuming image has a single channel).
    mask = thresh_gray > 0
    # Coordinates of non-black pixels.
    coords = np.argwhere(mask)
    if(coords.size > 0):
        # Bounding box of non-black pixels.
        x0, y0 = coords.min(axis=0)
        x1, y1 = coords.max(axis=0) #+ 1   # slices are exclusive at the top
        # Get the contents of the bounding box.
        cropped = colr_img[x0:x1, y0:y1]
        cropped = cv2.resize(cropped, (512, 512)) #(width, height)
```

```python
    #Smooth out the cropped fundus image
    scale = 300
    scaled = scaleRadius(cropped, scale)
    smoothed = cv2.addWeighted(scaled, 4, cv2.GaussianBlur(scaled , (0, 0), scale /30) ,-4, 128)
    #remove outer 10%,
    b = np.zeros(smoothed.shape[:2], np.uint8)
    cv2.circle(b, (smoothed.shape[1]//2, smoothed.shape[0]//2), int(scale*0.9), (1,1,1), -1,8,0)
    m= cv2.bitwise_and(smoothed, smoothed, mask=b)

    #Crop out black rounding pixels from smoothed images
    gray_img = cv2.cvtColor(m, cv2.COLOR_BGR2GRAY)
    retval, thresh_gray = cv2.threshold(gray_img, thresh=50, maxval=255,
type=cv2.THRESH_BINARY)
    mask = thresh_gray > 0
    coords = np.argwhere(mask)
    if(coords.size > 0):
       x0, y0 = coords.min(axis=0)
       x1, y1 = coords.max(axis=0)
       final_cropped = m[x0:x1, y0:y1]
    final_img = Image.fromarray(final_cropped)
    final_img.save(output_path)
```

## A.2  SOURCE CODE LISTINGS FOR TRANSFER LEARNING BASED MODEL TRAINING

```python
"""

Created on Thu Jun 10 02:19:11 2019

By Misgina Tsighe Hagos

"""

from PIL import Image
from matplotlib import pyplot as plt
# Load the Drive helper and mount
from google.colab import drive

# This will prompt for authorization.
drive.mount('/content/drive')

from keras import applications
from keras.preprocessing.image import ImageDataGenerator
from keras import optimizers
from keras.models import Sequential, Model
from keras.layers import Dropout, Flatten, Dense, GlobalAveragePooling2D, BatchNormalization, Activation
from keras import backend as k
from keras.callbacks import ModelCheckpoint, LearningRateScheduler, TensorBoard, EarlyStopping
from keras.models import model_from_yaml
from keras.preprocessing import image
import keras.backend as K
import tensorflow as tf
import os
import numpy as np
```

```python
import time
#import cv
from PIL import Image
from matplotlib import pyplot as plt
import math
from keras.callbacks import LearningRateScheduler, Callback

x = []
model_final = []
validation_data_dir=[]
train_data_dir=[]
train_data_dir.append("/content/drive/My Drive/Colab Notebooks/oaa_bin_class_500/sub_train_1vs1/")
validation_data_dir.append("/content/drive/My Drive/Colab Notebooks/five_classes_preed/validate/")

nb_train_samples = 600 #2600 #2710 #1300 pos #5000 #10 #60 #200 #4125
nb_validation_samples = 5000 #5007 #1000 #2000 #5000 #10 #60 #10 #466
batch_size = 50
epochs = 300
batch_size_validation = 2
img_width, img_height = 299, 299
classes = [["0","1"], ["0","2"], ["0","3"], ["0","4"]]
training_history = []

init_lr = 0.00001
def decay(epoch, steps=100):
    init_lr = 0.00001 #0.0003
    drop = 0.96
    epochs_drop = 8
    #lrate = init_lr * math.pow(drop, math.floor((1+epoch)/epochs_drop)) #Decaying
    lrate = init_lr / math.pow(drop, math.floor((1+epoch)/epochs_drop)) #Ascending
    if(lrate>0.0009): #if learning rate is above 0.01, keep it at it
        lrate = 0.0001
    return lrate

model = applications.inception_v3.InceptionV3(weights = "imagenet", include_top=False, input_shape = (img_width, img_height, 3)) #,classes=2
for layer in model.layers:
    layer.trainable = False

class EarlyStoppingByAccuracy(Callback):
    def __init__(self, monitor='accuracy', value=0.98, verbose=0):
        super(Callback, self).__init__()
        self.monitor = monitor
        self.value = value
        self.verbose = verbose
    def on_epoch_end(self, epoch, logs={}):
        current = logs.get(self.monitor)
        if current is None:
            warnings.warn("Early stopping requires %s available!" % self.monitor, RuntimeWarning)
        if current >= self.value:
            if self.verbose > 0:
                print("Epoch %05d: early stopping THR" % epoch)
            self.model.stop_training = True
```

```
i=0
print("In module: "+str(i))
x.insert(i, model.output)
x.insert(i, Flatten()(x[i]))
x.insert(i, Dense(32)(x[i])) #64 gave good results.
x.insert(i, Activation('relu')(x[i]))
x.insert(i, Dropout(0.5)(x[i])) #0.5
x.insert(i, BatchNormalization()(x[i]))
predictions = Dense(2, activation="softmax")(x[i]) #softmax

train_datagen = ImageDataGenerator(
rescale = 255./255)
test_datagen = ImageDataGenerator(
rescale = 255./255)

for j in range(len(classes)):
   c = classes[j]
   print(c)

   # creating and compiling the final model
   model_final.append(Model(inputs = model.input, outputs = predictions))
   model_final[j].compile(loss = "hinge", optimizer = optimizers.SGD(lr=init_lr, momentum=0.9, nesterov= True), metrics=["accuracy"])

   train_generator = train_datagen.flow_from_directory(
      train_data_dir[i],
      target_size = (img_width ,img_height),
      batch_size = batch_size,
      shuffle=True,
      class_mode = "categorical",
      classes = c)

   validation_generator = test_datagen.flow_from_directory(
      validation_data_dir[i],
      target_size = (img_width ,img_height),
      batch_size= batch_size_validation,
      shuffle=False,
      class_mode = "categorical",
      classes = ["0", "1"]) #  categorical binary

   train_generator.reset()
   validation_generator.reset()

   # Save the model according to the conditions
   checkpoint = ModelCheckpoint(str(j)+"_1vs1_model.h5", monitor='val_acc', verbose=1, save_best_only=True, save_weights_only=False, mode='auto', period=1)
   early = EarlyStopping(monitor='val_acc', min_delta=0, patience=40, verbose=1, mode='auto')

   label_map = (train_generator.class_indices)
   print(label_map)
   label_map = (validation_generator.class_indices)
   print(label_map)

   lr_sc = LearningRateScheduler(decay, verbose=1)
   callbacks = [
```

```
        EarlyStoppingByAccuracy(monitor='val_acc', value=0.95, verbose=1),  #Stop training if val_acc exceeds 88%
        lr_sc,
        checkpoint,
        early
    ]

    start_time = time.time()
    training_history.append(model_final[j].fit_generator(
    train_generator,
    steps_per_epoch = nb_train_samples//batch_size,
    epochs = epochs,
    callbacks=callbacks,
    validation_data = validation_generator,
    validation_steps = nb_validation_samples//batch_size_validation
    ))
```

## A.3 SOURCE CODE LISTINGS FOR SMALL INCEPTION NETWORK MODEL TRAINING

```
"""

Created on Thu Jun 14 05:30:22 2019

By Misgina Tsighe Hagos

"""

from PIL import Image
# Load the Drive helper and mount
from google.colab import drive

# This will prompt for authorization.
drive.mount('/content/drive')

import numpy as np
from keras.preprocessing.image import ImageDataGenerator
from keras.models import Sequential
from keras.layers import Dropout, Flatten, Dense, BatchNormalization, Activation, Input
from keras import applications, optimizers, regularizers
from keras.callbacks import ModelCheckpoint, LearningRateScheduler, TensorBoard, EarlyStopping, Callback
import time
from sklearn.utils import shuffle
#For building model from scratch
import keras
from keras.models import Model
from keras.layers import Conv2D, MaxPool2D, Dropout, Dense, Input, concatenate, GlobalAveragePooling2D, AveragePooling2D, Flatten
import math
from keras.utils import np_utils
from matplotlib import pyplot as plt

init_lr = 0.0001 #good for Adam 0.000001
def decay(epoch, steps=100):
    init_lr = 0.0001
```

```
    drop = 0.96
    epochs_drop = 7
    lrate = init_lr / math.pow(drop, math.floor((1+epoch)/epochs_drop)) #Ascending
    return lrate

train_data_dir = "/content/drive/My Drive/Colab Notebooks/fife_dr/train/"
#train_data_dir.append("/content/drive/My Drive/Colab Notebooks/five_classes_preed/train/")
validation_data_dir = "/content/drive/My Drive/Colab Notebooks/fife_dr/validate/"
model = []
training_history = []
nb_train_samples = 1000 #5000
nb_validation_samples = 1000
batch_size = 50
epochs = 400
batch_size_validation = 2
img_width, img_height = 200, 200
classes = [ ["0","1"], ["0","2"], ["0","3"], ["0","4"], ["1","2"],  ["1","3"], ["1","4"], ["2","3"], ["2","4"], ["3","4"]]

#%% Func to create inception module
def inception_module(x,
            filters_1x1,
            filters_3x3_reduce,
            filters_3x3,
            filters_5x5_reduce,
            filters_5x5,
            filters_pool_proj,
            name=None):

    conv_1x1 = Conv2D(filters_1x1, (1, 1), padding='same', activation='relu', kernel_initializer=kernel_init, bias_initializer=bias_init)(x)
    conv_1x1 = BatchNormalization()(conv_1x1)
    conv_3x3 = Conv2D(filters_3x3_reduce, (1, 1), padding='same', activation='relu', kernel_initializer=kernel_init, bias_initializer=bias_init)(x)
    conv_3x3 = Conv2D(filters_3x3, (3, 3), padding='same', activation='relu', kernel_initializer=kernel_init, bias_initializer=bias_init)(conv_3x3)
    conv_3x3 = BatchNormalization()(conv_3x3)
    conv_5x5 = Conv2D(filters_5x5_reduce, (1, 1), padding='same', activation='relu', kernel_initializer=kernel_init, bias_initializer=bias_init)(x)
    conv_5x5 = Conv2D(filters_5x5, (5, 5), padding='same', activation='relu', kernel_initializer=kernel_init, bias_initializer=bias_init)(conv_5x5)
    conv_5x5 = BatchNormalization()(conv_5x5)
    pool_proj = MaxPool2D((3, 3), strides=(1, 1), padding='same')(x)
    pool_proj = Conv2D(filters_pool_proj, (1, 1), padding='same', activation='relu', kernel_initializer=kernel_init, bias_initializer=bias_init)(pool_proj)
    output = concatenate([conv_1x1, conv_3x3, conv_5x5, pool_proj], axis=3, name=name)
    return output

kernel_init = keras.initializers.he_uniform() #he_uniform() #he_normal()  #glorot_uniform()
bias_init = keras.initializers.Constant(value=0.15)

#%% Define the small inception network
input_layer = Input(shape=(img_width, img_height, 3)) #Work with grayscale (3 to 1) images to reduce compute time
```

```
x = Conv2D(32, (5, 5), padding='same', strides=(2, 2), activation='relu', name='conv_1_5x5/2',
kernel_initializer=kernel_init, bias_initializer=bias_init)(input_layer)
x = BatchNormalization()(x)
x = MaxPool2D((3, 3), padding='same', strides=(2, 2), name='max_pool_1_3x3/2')(x)
x = Conv2D(64, (3, 3), padding='same', strides=(1, 1), activation='relu', name='conv_2a_3x3/1')(x)
x = BatchNormalization()(x)
x = MaxPool2D((3, 3), padding='same', strides=(2, 2), name='max_pool_2_3x3/2')(x)

x = inception_module(x,
            filters_1x1=64,
            filters_3x3_reduce=40,
            filters_3x3=60,
            filters_5x5_reduce=16,
            filters_5x5=32,
            filters_pool_proj=32,
            name='inception_3a')
x = MaxPool2D((3, 3), padding='same', strides=(2, 2), name='max_pool_3_3x3/1')(x)

x = inception_module(x,
            filters_1x1=60,
            filters_3x3_reduce=60,
            filters_3x3=80,
            filters_5x5_reduce=32,
            filters_5x5=64,
            filters_pool_proj=64,
            name='inception_3b')
x = GlobalAveragePooling2D(name='avg_pool_5_3x3/1')(x)
x = Dense(32, activation='relu', name='dense_final')(x)
x = BatchNormalization()(x)
x = Dropout(0.5)(x)
x = Dense(2, activation='softmax', name='output')(x) #5 total
model_v1_modified = Model(input_layer, x, name='inception_v1')

train_datagen = ImageDataGenerator(
#horizontal_flip=True,
#vertical_flip=True,
#rotation_range=90,
#fill_mode='wrap', #nearest
rescale = 255./255)

test_datagen = ImageDataGenerator(
rescale = 255./255)
print(len(classes))

for i in range(len(classes)):
   c = classes[i]
   print(c)
   train_generator = train_datagen.flow_from_directory(
      train_data_dir,
      target_size = (img_width, img_height),
      batch_size = batch_size,
      shuffle=True,
      class_mode = "categorical",
      classes = c)
   validation_generator = test_datagen.flow_from_directory(
```

```
        validation_data_dir,
        target_size = (img_width, img_height),
        batch_size= batch_size_validation,
        shuffle=False,
        class_mode = "categorical",
        classes = c)

    train_generator.reset()
    validation_generator.reset()

    #Start Model Training from here on
    #Assign the empty model_v1_modified to the current model
    model.append(model_v1_modified)
    lr_sc = LearningRateScheduler(decay, verbose=1)
    callbacks = [
        EarlyStoppingByAccuracy(monitor='val_acc', value=0.999, verbose=1),
        ModelCheckpoint(str(i)+"_bin_model.h5", monitor='val_acc', verbose=1, save_best_only=True,
save_weights_only=False, mode='auto', period=1),
        EarlyStopping(monitor='val_acc', min_delta=0, patience=50, verbose=1, mode='auto'),
        lr_sc
    ]

    model[i].compile(loss='hinge', optimizer=optimizers.SGD(lr=init_lr, momentum=0.9, nesterov=
True), metrics=['accuracy'])

    training_history.append(model[i].fit_generator(
    train_generator,
    steps_per_epoch = nb_train_samples//batch_size,
    epochs = epochs,
    callbacks = callbacks,
    validation_data = validation_generator,
    validation_steps = nb_validation_samples//batch_size_validation
    ))
```

## A.4  SOURCE CODE LISTINGS FOR MODEL ENSEMBLE

```
"""

Created on Thu Jul 18 01:09:10 2019

By Misgina Tsighe Hagos

"""

from keras.models import load_model
from keras.preprocessing import image
import numpy as np
import os

img_width,img_height = 200, 200
saved_models = ["0_bin_model.h5", "1_bin_model.h5", "2_bin_model.h5", "3_bin_model.h5"]
loaded_models = []
for i in range(4):
    loaded_models.append(load_model("D:/Edc/DS/Class/Modular
ANNs/Examples/oaa_bin_models/small 0VsAll in binary/" + saved_models[i]))
```

```python
pred_class = []

#%% Prediction and Aggregation
folder_path = 'C:/Users/Misgun/Documents/Python Scripts/fully_graham_preped/4_preped/'

for img in os.listdir(folder_path):
    imagePath = os.path.join(folder_path, img)
    test_image = image.load_img(imagePath, target_size = (img_width, img_height))
    test_image = image.img_to_array(test_image)
    test_image = np.expand_dims(test_image, axis = 0)
    
    #predict the result
    result = loaded_models[3].predict(test_image)
    labels = np.argmax(result,axis=1)
    if(labels==1):
        print("Proliferative Classifier")
        print(result)
        print(labels)
        pred_class.append(labels)
        print()
        continue
    else:
        result = loaded_models[2].predict(test_image)
        labels = np.argmax(result,axis=1)
        if(labels==1):
            print("Severe Classifier")
            print(result)
            print(labels)
            pred_class.append(labels)
            print()
            continue
        else:
            result = loaded_models[1].predict(test_image)
            labels = np.argmax(result,axis=1)
            if(labels==1):
                print("Moderate Classifier")
                print(result)
                print(labels)
                pred_class.append(labels)
                print()
                continue
            else:
                result = loaded_models[0].predict(test_image)
                labels = np.argmax(result,axis=1)
                if(labels==1):
                    print("Mild Classifier")
                    print(result)
                    print(labels)
                    pred_class.append(labels)
                    print()
                    continue
                else:
                    print("Normal")
                    print(result)
```

```
            print(labels)
            pred_class.append(labels)
            print()
sum(x==1 for x in pred_class)
```

# REFERENCES

[1] Yau, J.W., Rogers, S.L., Kawasaki, R., Lamoureux, E.L., Kowalski, J.W., Bek, T., Chen, S.J., Dekker, J.M., Fletcher, A., Grauslund, J. and Haffner, S., 2012. Global prevalence and major risk factors of diabetic retinopathy. *Diabetes care*, *35*(3), pp.556-564.

[2] Boyle, J.P., Honeycutt, A.A., Narayan, K.V., Hoerger, T.J., Geiss, L.S., Chen, H. and Thompson, T.J., 2001. Projection of diabetes burden through 2050: impact of changing demography and disease prevalence in the US. *Diabetes care*, *24*(11), pp.1936-1940.

[3] Leasher, J.L., Bourne, R.R., Flaxman, S.R., Jonas, J.B., Keeffe, J., Naidoo, K., Pesudovs, K., Price, H., White, R.A., Wong, T.Y. and Resnikoff, S., 2016. Global estimates on the number of people blind or visually impaired by diabetic retinopathy: a meta-analysis from 1990 to 2010. *Diabetes care*, 39(9), pp.1643-1649.

[4] "Diabetic Retinopathy", *National Eye Institute*, 2019. [Online]. Available: https://nei.nih.gov/eyedata/diabetic. [Accessed: 12- Jul- 2019].

[5] Wilkinson, C.P., Ferris III, F.L., Klein, R.E., Lee, P.P., Agardh, C.D., Davis, M., Dills, D., Kampik, A., Pararajasegaram, R., Verdaguer, J.T. and Group, G.D.R.P., 2003. Proposed international clinical diabetic retinopathy and diabetic macular edema disease severity scales. *Ophthalmology*, *110*(9), pp.1677-1682.

[6] Salz, D.A. and Witkin, A.J., 2015. Imaging in diabetic retinopathy. *Middle East African journal of ophthalmology*, *22*(2), p.145.

[7] Mookiah, M.R.K., Acharya, U.R., Chua, C.K., Lim, C.M., Ng, E.Y.K. and Laude, A., 2013. Computer-aided diagnosis of diabetic retinopathy: A review. *Computers in biology and medicine*, 43(12), pp.2136-2155.

[8] Foster, A. and Resnikoff, S., 2005. The impact of Vision 2020 on global blindness. *Eye*, *19*(10), p.1133.

[9] Kim TN, Myers F, Reber C, Loury PJ, Loumou P, Webster D, et al. A smartphone-based tool for rapid, portable, and automated wide-field retinal imaging. Transl Vis Sci Technol. 2018;7:21.

[10] LeCun, Y., Boser, B.E., Denker, J.S., Henderson, D., Howard, R.E., Hubbard, W.E. and Jackel, L.D., 1990. Handwritten digit recognition with a back-propagation network. In *Advances in neural information processing systems* (pp. 396-404).

[11] Miotto, R., Wang, F., Wang, S., Jiang, X. and Dudley, J.T., 2017. Deep learning for healthcare: review, opportunities and challenges. *Briefings in bioinformatics*, *19*(6), pp.1236-1246.

[12] Razzak, M.I., Naz, S. and Zaib, A., 2018. Deep learning for medical image processing: Overview, challenges and the future. In *Classification in BioApps* (pp. 323-350). Springer, Cham.

[13] Erhan, D., Bengio, Y., Courville, A., Manzagol, P.A., Vincent, P. and Bengio, S., 2010. Why does unsupervised pre-training help deep learning? *Journal of Machine Learning Research*, *11*(Feb), pp.625-660.

[14] Litjens, G., Kooi, T., Bejnordi, B.E., Setio, A.A.A., Ciompi, F., Ghafoorian, M., Van Der Laak, J.A., Van Ginneken, B. and Sánchez, C.I., 2017. A survey on deep learning in medical image analysis. *Medical image analysis*, *42*, pp.60-88.

[15] Altaf, F., Islam, S., Akhtar, N. and Janjua, N.K., 2019. Going Deep in Medical Image Analysis: Concepts, Methods, Challenges and Future Directions. *arXiv preprint arXiv:1902.05655*.

[16] Deng, J., Dong, W., Socher, R., Li, L.J., Li, K. and Fei-Fei, L., 2009, June. Imagenet: A large-scale hierarchical image database. In *2009 IEEE conference on computer vision and pattern recognition* (pp. 248-255). Ieee.

[17]     Chollet, F., 2017. Xception: Deep learning with depthwise separable convolutions. In *Proceedings of the IEEE conference on computer vision and pattern recognition* (pp. 1251-1258).

[18]     Simonyan, K. and Zisserman, A., 2014. Very deep convolutional networks for large-scale image recognition. *arXiv preprint arXiv:1409.1556*.

[19]     He, K., Zhang, X., Ren, S. and Sun, J., 2016. Deep residual learning for image recognition. In *Proceedings of the IEEE conference on computer vision and pattern recognition* (pp. 770-778).

[20]     He, K., Zhang, X., Ren, S. and Sun, J., 2016, October. Identity mappings in deep residual networks. In *European conference on computer vision* (pp. 630-645). Springer, Cham.

[21]     Xie, S., Girshick, R., Dollár, P., Tu, Z. and He, K., 2017. Aggregated residual transformations for deep neural networks. In *Proceedings of the IEEE conference on computer vision and pattern recognition* (pp. 1492-1500).

[22]     Szegedy, C., Vanhoucke, V., Ioffe, S., Shlens, J. and Wojna, Z., 2016. Rethinking the inception architecture for computer vision. In *Proceedings of the IEEE conference on computer vision and pattern recognition* (pp. 2818-2826).

[23]     Szegedy, C., Ioffe, S., Vanhoucke, V. and Alemi, A.A., 2017, February. Inception-v4, inception-resnet and the impact of residual connections on learning. In *Thirty-First AAAI Conference on Artificial Intelligence*.

[24]     Howard, A.G., Zhu, M., Chen, B., Kalenichenko, D., Wang, W., Weyand, T., Andreetto, M. and Adam, H., 2017. Mobilenets: Efficient convolutional neural networks for mobile vision applications. *arXiv preprint arXiv:1704.04861*.

[25]     Sandler, M., Howard, A., Zhu, M., Zhmoginov, A. and Chen, L.C., 2018. Mobilenetv2: Inverted residuals and linear bottlenecks. In *Proceedings of the IEEE Conference on Computer Vision and Pattern Recognition* (pp. 4510-4520).

[26]     Huang, G., Liu, Z., Van Der Maaten, L. and Weinberger, K.Q., 2017. Densely connected convolutional networks. In *Proceedings of the IEEE conference on computer vision and pattern recognition* (pp. 4700-4708).

[27]     Zoph, B., Vasudevan, V., Shlens, J. and Le, Q.V., 2018. Learning transferable architectures for scalable image recognition. In *Proceedings of the IEEE conference on computer vision and pattern recognition* (pp. 8697-8710).

[28]     Colaboratory. (2018). *Welcome to Colaboratory*. [online] Available at: http://colab.research.google.com/ [Accessed 2 May 2019].

[29]     Carneiro, T., Da Nóbrega, R.V.M., Nepomuceno, T., Bian, G.B., De Albuquerque, V.H.C. and Reboucas Filho, P.P., 2018. Performance Analysis of Google Colaboratory as a Tool for Accelerating Deep Learning Applications. *IEEE Access*, *6*, pp.61677-61685.

[30]     Kaggle. (2015). *Diabetic Retinopathy Detection*. [online] Available at: https://www.kaggle.com/c/diabetic-retinopathy-detection [Accessed 11 Mar. 2019].

[31]     Decencière, E., Zhang, X., Cazuguel, G., Lay, B., Cochener, B., Trone, C., Gain, P., Ordonez, R., Massin, P., Erginay, A. and Charton, B., 2014. Feedback on a publicly distributed image database: the Messidor database. *Image Analysis & Stereology*, *33*(3), pp.231-234.

[32]     E-ophtha. [online] Available at: http://www.adcis.net/en/Download-Third-Party/E-Ophtha.html [Accessed: 11 Mar. 2019]

[33]     Drive dataset. [online] Available at: https://www.isi.uu.nl/Research/Databases/DRIVE/ [Accessed: 11 Mar 2019]

[34]     Hoover, A., Kouznetsova, V. and Goldbaum, M., 1998. Locating blood vessels in retinal images by piece-wise threshold probing of a matched filter response. In

[34] *Proceedings of the AMIA Symposium* (p. 931). American Medical Informatics Association.

[35] Kälviäinen, R.V.J.P.H. and Uusitalo, H., 2007. DIARETDB1 diabetic retinopathy database and evaluation protocol. In *Medical Image Understanding and Analysis* (Vol. 2007, p. 61).

[36] Diaretdb1 dataset. [online] Available at: http://www.it.lut.fi/project/imageret/diaretdb1/ [Accessed: 11 Mar 2019]

[37] Chase dataset. [online] Available at: http://www.chasestudy.ac.uk/ [Accessed: 11 Mar 2019]

[38] Prentašić, P., Lončarić, S., Vatavuk, Z., Benčić, G., Subašić, M., Petković, T., Dujmović, L., Malenica-Ravlić, M., Budimlija, N. and Tadić, R., 2013, September. Diabetic retinopathy image database (DRiDB): a new database for diabetic retinopathy screening programs research. In *2013 8th International Symposium on Image and Signal Processing and Analysis (ISPA)* (pp. 711-716). IEEE.

[39] Zhang, Z., Yin, F.S., Liu, J., Wong, W.K., Tan, N.M., Lee, B.H., Cheng, J. and Wong, T.Y., 2010, August. Origa-light: An online retinal fundus image database for glaucoma analysis and research. In *2010 Annual International Conference of the IEEE Engineering in Medicine and Biology* (pp. 3065-3068). IEEE.

[40] Sng, C.C., Foo, L.L., Cheng, C.Y., Allen Jr, J.C., He, M., Krishnaswamy, G., Nongpiur, M.E., Friedman, D.S., Wong, T.Y. and Aung, T., 2012. Determinants of anterior chamber depth: the Singapore Chinese Eye Study. *Ophthalmology*, *119*(6), pp.1143-1150.

[41] Nih areds dataset. [online] Available at: https://www.nih.gov/news-events/news-releases/nih-adds-first-images-major-research-database [Accessed: 11 Mar 2019]

[42] Al-Diri, B., Hunter, A., Steel, D., Habib, M., Hudaib, T. and Berry, S., 2008, August. A reference data set for retinal vessel profiles. In *2008 30th Annual International Conference of the IEEE Engineering in Medicine and Biology Society* (pp. 2262-2265). IEEE.

[43] Eyepacs dataset. [online] Available at: http://www.eyepacs.com/eyepacssystem/ [Accessed: 11 Mar 2019]

[44] Fumero, F., Alayón, S., Sanchez, J.L., Sigut, J. and Gonzalez-Hernandez, M., 2011, June. RIM-ONE: An open retinal image database for optic nerve evaluation. In *2011 24th international symposium on computer-based medical systems (CBMS)* (pp. 1-6). IEEE.

[45] Sivaswamy, J., Krishnadas, S.R., Joshi, G.D., Jain, M. and Tabish, A.U.S., 2014, April. Drishti-gs: Retinal image dataset for optic nerve head (onh) segmentation. In *2014 IEEE 11th international symposium on biomedical imaging (ISBI)* (pp. 53-56). IEEE.

[46] Aria dataset. [online] Available at: http://www.eyecharity.com/aria_online.html. [Accessed: 11 Mar 2019]

[47] drion dataset. [online] Available at: http://www.ia.uned.es/~ejcarmona/DRIONS-DB.html [Accessed: 11 Mar 2019]

[48] Seed-db. [online] Available at: https://www.seri.com.sg/key-programmes/singapore-epidemiology-of-eye-diseases-seed/ [Accessed: 11 Mar 2019]

[49] Abràmoff, M.D., Lou, Y., Erginay, A., Clarida, W., Amelon, R., Folk, J.C. and Niemeijer, M., 2016. Improved automated detection of diabetic retinopathy on a publicly available dataset through integration of deep learning. *Investigative ophthalmology & visual science*, *57*(13), pp.5200-5206.

[50] Mo, J., Zhang, L. and Feng, Y., 2018. Exudate-based diabetic macular edema recognition in retinal images using cascaded deep residual networks. *Neurocomputing*, *290*, pp.161-171.

[51] Perdomo, O., Otalora, S., Rodríguez, F., Arevalo, J. and González, F.A., 2016. A novel machine learning model based on exudate localization to detect diabetic macular edema.

[52] Burlina, P., Freund, D.E., Joshi, N., Wolfson, Y. and Bressler, N.M., 2016, April. Detection of age-related macular degeneration via deep learning. In 2016 *IEEE 13th International Symposium on Biomedical Imaging* (ISBI) (pp. 184-188). IEEE.

[53] Al-Bander, B., Al-Nuaimy, W., Al-Taee, M.A., Williams, B.M. and Zheng, Y., 2016. Diabetic macular edema grading based on deep neural networks.

[54] Ting, D.S.W., Cheung, C.Y.L., Lim, G., Tan, G.S.W., Quang, N.D., Gan, A., Hamzah, H., Garcia-Franco, R., San Yeo, I.Y., Lee, S.Y. and Wong, E.Y.M., 2017. Development and validation of a deep learning system for diabetic retinopathy and related eye diseases using retinal images from multiethnic populations with diabetes. *Jama*, 318(22), pp.2211-2223.

[55] Arunkumar, R. and Karthigaikumar, P., 2017. Multi-retinal disease classification by reduced deep learning features. *Neural Computing and Applications*, 28(2), pp.329-334.

[56] Prentašić, P. and Lončarić, S., 2016. Detection of exudates in fundus photographs using deep neural networks and anatomical landmark detection fusion. *Computer methods and programs in biomedicine*, 137, pp.281-292.

[57] Perdomo, O., Arevalo, J. and González, F.A., 2017, January. Convolutional network to detect exudates in eye fundus images of diabetic subjects. In *12th International Symposium on Medical Information Processing and Analysis* (Vol. 10160, p. 101600T). International Society for Optics and Photonics.

[58] Gondal, W.M., Köhler, J.M., Grzeszick, R., Fink, G.A. and Hirsch, M., 2017, September. Weakly-supervised localization of diabetic retinopathy lesions in retinal fundus images. In *2017 IEEE International Conference on Image Processing (ICIP)* (pp. 2069-2073). IEEE.

[59] Quellec, G., Charrière, K., Boudi, Y., Cochener, B. and Lamard, M., 2017. Deep image mining for diabetic retinopathy screening. *Medical image analysis*, 39, pp.178-193.

[60] Haloi, M., 2015. Improved microaneurysm detection using deep neural networks. *arXiv preprint arXiv:1505.04424*.

[61] van Grinsven, M.J., van Ginneken, B., Hoyng, C.B., Theelen, T. and Sánchez, C.I., 2016. Fast convolutional neural network training using selective data sampling: Application to hemorrhage detection in color fundus images. *IEEE transactions on medical imaging*, 35(5), pp.1273-1284.

[62] Orlando, J.I., Prokofyeva, E., del Fresno, M. and Blaschko, M.B., 2018. An ensemble deep learning based approach for red lesion detection in fundus images. *Computer methods and programs in biomedicine*, 153, pp.115-127.

[63] Shan, J. and Li, L., 2016, June. A deep learning method for microaneurysm detection in fundus images. In *2016 IEEE First International Conference on Connected Health: Applications, Systems and Engineering Technologies (CHASE)* (pp. 357-358). IEEE.

[64] Li, T., Gao, Y., Wang, K., Guo, S., Liu, H. and Kang, H., 2019. Diagnostic Assessment of Deep Learning Algorithms for Diabetic Retinopathy Screening. *Information Sciences*.

[65] DDR-dataset. online Available at: **https://github.com/nkicsl/DDR-dataset** [Accessed: 11 Mar 2019]

[66] Nagasawa, T., Tabuchi, H., Masumoto, H., Enno, H., Niki, M., Ohara, Z., Yoshizumi, Y., Ohsugi, H. and Mitamura, Y., 2019. Accuracy of ultrawide-field fundus ophthalmoscopy-assisted deep learning for detecting treatment-naïve proliferative diabetic retinopathy. *International ophthalmology*, pp.1-7.

[67] Zeng, X., Chen, H., Luo, Y. and Ye, W., 2019. Automated Diabetic Retinopathy Detection Based on Binocular Siamese-Like Convolutional Neural Network. *IEEE Access*, *7*, pp.30744-30753.

[68] Zhao, Z., Zhang, K., Hao, X., Tian, J., Chua, M.C.H., Chen, L. and Xu, X., 2019. BiRA-Net: Bilinear Attention Net for Diabetic Retinopathy Grading. *arXiv preprint arXiv:1905.06312*.

[69] Verbraak, F.D., Abramoff, M.D., Bausch, G.C., Klaver, C., Nijpels, G., Schlingemann, R.O. and van der Heijden, A.A., 2019. Diagnostic Accuracy of a Device for the Automated Detection of Diabetic Retinopathy in a Primary Care Setting. *Diabetes care*, *42*(4), pp.651-656.

[70] Abràmoff, M.D., Lavin, P.T., Birch, M., Shah, N. and Folk, J.C., 2018. Pivotal trial of an autonomous AI-based diagnostic system for detection of diabetic retinopathy in primary care offices. *Npj Digital Medicine*, *1*(1), p.39.

[71] González-Gonzalo, C., Sánchez-Gutiérrez, V., Hernández-Martínez, P., Contreras, I., Lechanteur, Y.T., Domanian, A., van Ginneken, B. and Sánchez, C.I., 2019. Evaluation of a deep learning system for the joint automated detection of diabetic retinopathy and age-related macular degeneration. *arXiv preprint arXiv:1903.09555*.

[72] Hagos, M.T. and Kant, S., 2019. Transfer Learning based Detection of Diabetic Retinopathy from Small Dataset. *arXiv preprint arXiv:1905.07203*.

[73] Gulshan, V., Peng, L., Coram, M., Stumpe, M.C., Wu, D., Narayanaswamy, A., Venugopalan, S., Widner, K., Madams, T., Cuadros, J. and Kim, R., 2016. Development and validation of a deep learning algorithm for detection of diabetic retinopathy in retinal fundus photographs. *Jama*, 316(22), pp.2402-2410.

[74] Colas, E., Besse, A., Orgogozo, A., Schmauch, B., Meric, N. and Besse, E., 2016. Deep learning approach for diabetic retinopathy screening. *Acta Ophthalmologica*, 94.

[75] Costa, P. and Campilho, A., 2017. Convolutional bag of words for diabetic retinopathy detection from eye fundus images. *IPSJ Transactions on Computer Vision and Applications*, 9(1), p.10.

[76] Pratt, H., Coenen, F., Broadbent, D.M., Harding, S.P. and Zheng, Y., 2016. Convolutional neural networks for diabetic retinopathy. *Procedia Computer Science*, 90, pp.200-205.

[77] Gargeya, R. and Leng, T., 2017. Automated identification of diabetic retinopathy using deep learning. *Ophthalmology*, 124(7), pp.962-969.

[78] Mansour, R.F., 2018. Deep-learning-based automatic computer-aided diagnosis system for diabetic retinopathy. *Biomedical engineering letters*, 8(1), pp.41-57.

[79] Cavallerano, J., Lawrence, M.G., Zimmer-Galler, I., Bauman, W., Bursell, S., Gardner, W.K., Horton, M., Hildebrand, L., Federman, J., Carnahan, L. and Kuzmak, P., 2004. Telehealth practice recommendations for diabetic retinopathy. *Telemedicine journal and e-health: the official journal of the American Telemedicine Association*, *10*(4), pp.469-482.

[80] Bursell, S.E., Brazionis, L. and Jenkins, A., 2012. Telemedicine and ocular health in diabetes mellitus. *Clinical and Experimental Optometry*, *95*(3), pp.311-327.

[81] Coronado, A.C., 2014. Diagnostic accuracy of tele-ophthalmology for diabetic retinopathy assessment: A meta-analysis and economic analysis.

[82] Mohammadpour, M., Heidari, Z., Mirghorbani, M. and Hashemi, H., 2017. Smartphones, tele-ophthalmology, and VISION 2020. *International journal of ophthalmology*, *10*(12), p.1909.

[83] Prasanna, P., Jain, S., Bhagat, N. and Madabhushi, A., 2013, May. Decision support system for detection of diabetic retinopathy using smartphones. In *2013 7th International Conference on Pervasive Computing Technologies for Healthcare and Workshops* (pp. 176-179). IEEE.

[84] Xu, X., Ding, W., Wang, X., Cao, R., Zhang, M., Lv, P. and Xu, F., 2016. Smartphone-based accurate analysis of retinal vasculature towards point-of-care diagnostics. *Scientific reports*, *6*, p.34603.

[85] Rajalakshmi, R., Subashini, R., Anjana, R.M. and Mohan, V., 2018. Automated diabetic retinopathy detection in smartphone-based fundus photography using artificial intelligence. *Eye*, *32*(6), p.1138.

[86] Jamil, A.Z., Ali, L., Tahir, M.Y. and Shiraz, F.S., 2018. Smart Phone: A Smart Technology for Fundus Photography in Diabetic Retinopathy Screening. *Pakistan Journal of Ophthalmology*, *34*(4).

[87] Kashyap, N., Singh, D.K. and Singh, G.K., 2017, October. Mobile phone based diabetic retinopathy detection system using ANN-DWT. In *2017 4th IEEE Uttar Pradesh Section International Conference on Electrical, Computer and Electronics (UPCON)* (pp. 463-467). IEEE.

[88] Kashyap, N., Singh, D.K. and Singh, G.K., 2019. Color Histogram- and Smartphone-Based Diabetic Retinopathy Detection System. In *Engineering Vibration, Communication and Information Processing* (pp. 669-678). Springer, Singapore.

[89] Graham, B., 2015. Kaggle diabetic retinopathy detection competition report. *University of Warwick*.

[90] Barz, B. and Denzler, J., 2019. Deep Learning on Small Datasets without Pre-Training using Cosine Loss. *arXiv preprint arXiv:1901.09054*.

[91] Mohammadian, S., Karsaz, A. and Roshan, Y.M., 2017, November. Comparative Study of Fine-Tuning of Pre-Trained Convolutional Neural Networks for Diabetic Retinopathy Screening. In *2017 24th National and 2nd International Iranian Conference on Biomedical Engineering (ICBME)* (pp. 1-6). IEEE.